\documentclass[11pt,a4paper]{article}
\pdfoutput=1
\usepackage{jheppub_kim}

\usepackage{multirow, graphicx,amssymb,url,mathrsfs,amsmath}
\usepackage{wrapfig,boxedminipage,subfigure,epsfig}
\usepackage{amsxtra,amstext,latexsym,dsfont,amsfonts}
\usepackage{color}
\usepackage[dvipsnames]{xcolor}
\usepackage{float}
\usepackage{slashed}

\newcommand{\dd}{\mathrm{d}}

\newcommand{\expo}{{ e^{\frac{2 \pi (x+i \tau)}{\beta}}  }}
\newcommand{\expoo}{{ e^{\frac{2 \pi (x-i \tau)}{\beta}}  }}

\title{Entanglement and R\'{e}nyi Entropy of Multiple Intervals in  $T\overline{T}$-Deformed CFT and Holography}

\author{Hyun-Sik Jeong,}
\author{Keun-Young Kim,}
\author{and Mitsuhiro Nishida}

\emailAdd{hyunsik@gist.ac.kr}
\emailAdd{fortoe@gist.ac.kr}
\emailAdd{mnishida@gist.ac.kr}

\affiliation{ School of Physics and Chemistry, Gwangju Institute of Science and Technology, \\
123 Cheomdan-gwagiro, Gwangju 61005, Korea}

\abstract{
We study the entanglement entropy (EE) and the R\'{e}nyi entropy (RE) of multiple intervals in two-dimensional $T\overline{T}$-{\it deformed} conformal field theory (CFT) at {\it finite temperature} by field theoretic and holographic methods.  
First, by the replica method with the twist operators, we construct the general formula of the RE and EE up to the first order of a deformation parameter. 
By using our general formula, we show that the EE of multiple intervals for a holographic CFT is just a summation of the single interval case even with the small deformation. This is a non-trivial consequence from the field theory perspective, though it may be expected by the Ryu-Takayanagi formula in holography. 
However, the deformed RE of the two intervals is a summation of the single interval case only if the separations between the intervals are big enough. 
It can be understood by the tension of the cosmic branes dual to the RE.
We also study the holographic EE for single and two intervals with an {\it arbitrary} cut-off radius (dual to the $T\overline{T}$ deformation) at {\it any} temperature. We confirm our holographic results agree with the field theory results with a {\it small} deformation and {\it high} temperature limit, as expected. For two intervals, there are two configurations for EE: disconnected ($s$-channel) and connected ($t$-channel) ones. We investigate the phase transition between them as we change parameters: as the deformation or temperature increases the phase transition is suppressed and the disconnected phase is more favored. 
}

\begin{document}

\maketitle

\section{Introduction}
The AdS/CFT correspondence \cite{Maldacena:1997re, Gubser:1998bc, Witten:1998qj} is a mysterious duality between field theories and gravity theories. It gives a new geometric interpretation to a special class of field theories. To check this correspondence, the conformal symmetry often plays an important role in explicit computations of the partition function and correlation functions. It is an interesting problem to make computable examples for generalizations of the AdS/CFT correspondence by deforming the conformal symmetry. The authors of \cite{McGough:2016lol} proposed such an example: the AdS$_3$/CFT$_2$ correspondence with the $T\overline{T}$ deformation and a finite radius cutoff.

Let us briefly review the $T\overline{T}$ deformation of  2d CFT on a flat spacetime \cite{Zamolodchikov:2004ce,Smirnov:2016lqw, Cavaglia:2016oda}. Consider a deformed CFT by the $T\overline{T}$ operator with a deformation parameter $\mu$. The action ($S^{(\mu)}_{\textrm{QFT}}$) of this deformed CFT is defined via the differential equation
\begin{align}
\frac{d S^{(\mu)}_{\textrm{QFT}}}{d \mu}=\int \dd^2 x \,(T\overline{T})_{\mu}, \;\;\;\;S^{(\mu)}_{\textrm{QFT}}\Bigr|_{\mu=0}=S_\textrm{CFT},
\end{align}
where $S_\textrm{CFT}$ is the action of the undeformed CFT, and $(T\overline{T})_{\mu}$ is a local operator which is constructed by the energy momentum tensor of the deformed CFT. {See Eq. (6.10) in \cite{Cavaglia:2016oda} for a simple example.}

If we consider the first order perturbation in $\mu$, the perturbative action ($S_{\textrm{QFT}}$) is given by
\begin{align}
S_{\textrm{QFT}}=S_\textrm{CFT}+\mu \int \dd^2 x \, T\overline{T}\,,\label{pa}
\end{align}
where
\begin{align}
T:=T_{ww}\,, \qquad \overline{T}:=T_{\bar{w}\bar{w}} \,,
\end{align}
are the energy momentum tensors of the undeformed CFT and $w$ and $\bar{w}$ are complex coordinates.\footnote{We omit the $\Theta^2:=T_{w\bar{w}}T_{w\bar{w}}$ term because the correlation functions which include $\Theta^2$ on the cylinder are zero.} 
The $T\overline{T}$ deformation of 2d CFT has a  solvable structure. In particular, the energy spectrum on a spatial circle can be computed non-perturbatively \cite{Smirnov:2016lqw, Cavaglia:2016oda}. For recent studies on the $T\overline{T}$ deformation see, for example, \cite{Cardy:2015xaa, Giribet:2017imm, Dubovsky:2017cnj, Cardy:2018sdv, Aharony:2018vux, Bonelli:2018kik, Dubovsky:2018bmo,  Datta:2018thy, Conti:2018jho,  Chen:2018keo,      Aharony:2018bad,  Conti:2018tca, Santilli:2018xux, Baggio:2018rpv, Chang:2018dge, Jiang:2019tcq, LeFloch:2019rut, Jiang:2019hux, Conti:2019dxg, Chang:2019kiu, Frolov:2019nrr}.  Non-Lorentz invariant cases are also studied in, for example, \cite{Guica:2017lia, Chakraborty:2018vja, Aharony:2018ics, Cardy:2018jho, Nakayama:2018ujt,  Guica:2019vnb}.

The proposal in \cite{McGough:2016lol} is that the gravity dual of the $T\overline{T}$-deformed  2d holographic CFT with $\mu>0$ is AdS$_3$ gravity with a finite radius cutoff. This proposal has been studied and checked by various methods. In particular, the energy spectrum of the deformed CFT is  matched  to the quasi-local energy in the cutoff space time $r\le r_c$ \cite{McGough:2016lol}. See also recent development of this holography in \cite{Shyam:2017znq, Kraus:2018xrn, Cottrell:2018skz, Bzowski:2018pcy, Taylor:2018xcy, Hartman:2018tkw, Shyam:2018sro, Caputa:2019pam}. Another gravity dual of the deformed CFT for vacua of string theory  was proposed in \cite{Giveon:2017nie}, see also \cite{Giveon:2017myj, Asrat:2017tzd, Baggio:2018gct, Apolo:2018qpq, Babaro:2018cmq, Chakraborty:2018aji, Araujo:2018rho,  Giveon:2019fgr, Chakraborty:2019mdf, Nakayama:2019mvq, Dei:2018mfl, Dei:2018jyj}.

Holographic entanglement entropy \cite{Ryu:2006bv, Ryu:2006ef} is a well studied topic in the AdS/CFT correspondence. The entanglement entropy in the $T\overline{T}$-deformed CFT and its holographic dual have been studied in \cite{Chakraborty:2018kpr, Donnelly:2018bef, Chen:2018eqk, Gorbenko:2018oov, Park:2018snf, Sun:2019ijq, Banerjee:2019ewu, Murdia:2019fax, Ota:2019yfe}. 
Especially, a perturbative computation of the entanglement entropy in the $T\overline{T}$-deformed 2d CFT on a cylinder \cite{Chen:2018eqk} and its non-perturbative computation with a large central charge on a  sphere \cite{Donnelly:2018bef} are consistent with the holographic entanglement entropy with a radius cutoff.

Without the $T\overline{T}$ deformation, the entanglement entropy of a single interval in the ground state of 2d CFT is expressed by a well known formula \cite{Calabrese:2004eu, Calabrese:2009qy}. Even though it has been reproduced by holography, we note that it is valid in {\it any} CFT not only in the {\it holographic CFT}, {where the conditions for the holographic CFT are a large central charge and sparse spectrum.}
This universality follows from the universality of two point functions in CFT. 

On the other hand, the entanglement entropy of multiple intervals is related to higher point functions of the twist operators, which depend on the details of CFT. Thus, in order to check the AdS$_3$/CFT$_2$ correspondence for the entanglement entropy of multiple intervals, we need to use conditions for the {\it holographic CFT}. In \cite{Hartman:2013mia}, the entanglement entropy of multiple intervals in the 2d holographic CFT was computed by the dominant contribution (vacuum conformal block) in the correlation functions, and it  agrees with the holographic entanglement entropy formula. 
This agreement is an important consistency check for the holographic entanglement entropy formula because it is confirmed under the conditions of {\it holographic CFT}  in the field theory side.

With the first order $T\overline{T}$ deformation, the  R\'{e}nyi  entropy of a {\it single} interval in the deformed 2d free fermions CFT was studied at {\it zero} temperature  \cite{Chakraborty:2018kpr, Sun:2019ijq} {by using the twist operators. 
In this paper, we develop a formula for R\'{e}nyi entanglement entropy of {\it multiple} intervals in 2d CFT with the first order $T\overline{T}$ deformation at {\it finite} temperature by using the twist operators. Our formula reproduces the R\'{e}nyi entropy of a {\it single} interval in the deformed 2d CFT at finite temperature~\cite{ Chen:2018eqk}, where a different method, a conformal map from a replica manifold to a complex plane, was used. }

We find that the entanglement entropy of multiple intervals in the deformed holographic CFT is a summation of the one of the single interval because of the dominant contribution from the vacuum conformal block. 
The R\'{e}nyi  entropy of two intervals becomes a summation of the one of the single interval {if} the distance between the intervals is large enough. These `additivity' properties from the field theory are consistent with the holographic computation with a radius cutoff. 

\begin{figure}[]
\centering
     \subfigure[$s$-channel]
     {\includegraphics[width=7.3cm]{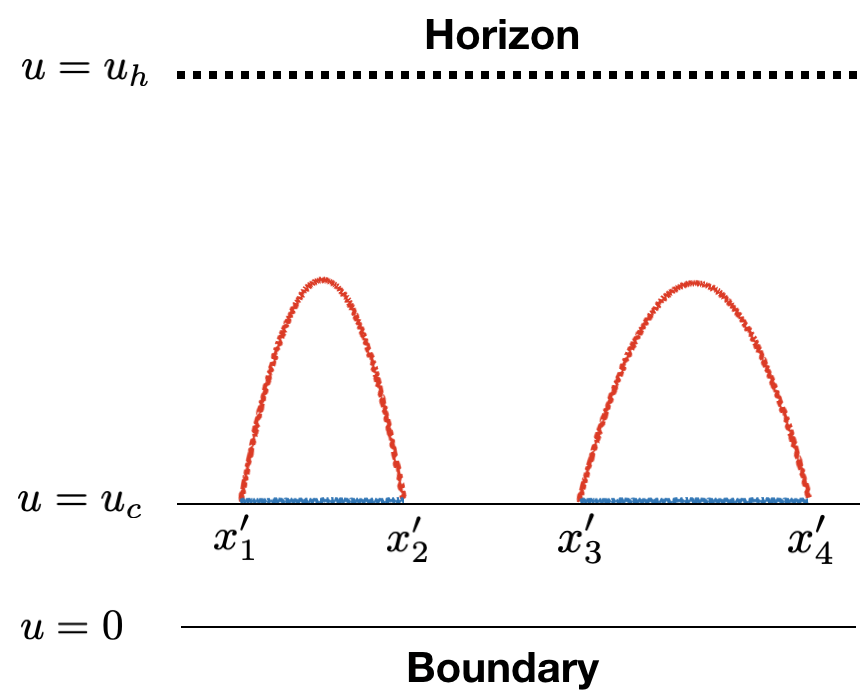} \label{}}
      \subfigure[$t$-channel]
     {\includegraphics[width=7.3cm]{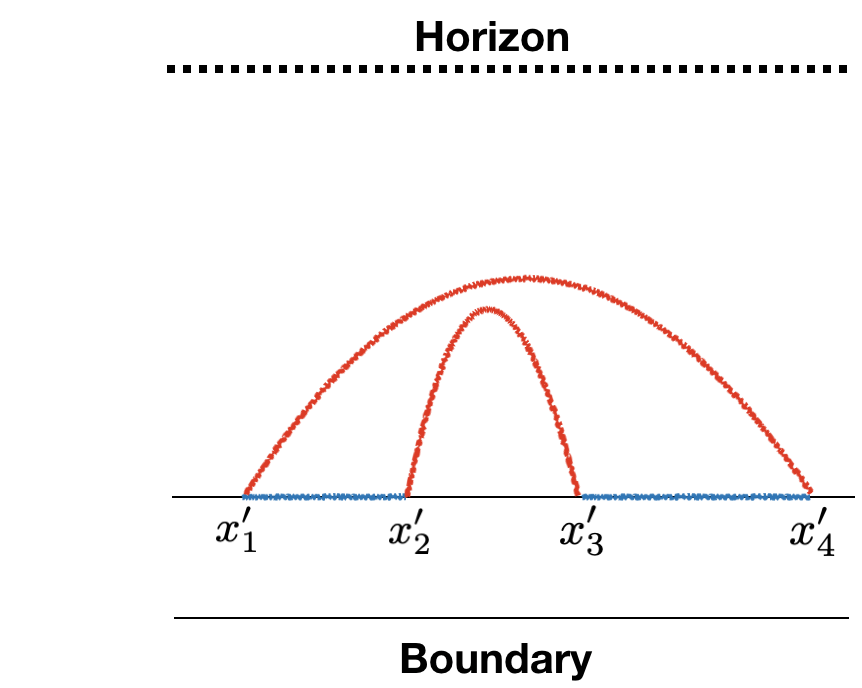} \label{}}
 \caption{The holographic entanglement entropy for two intervals $[x'_1, x'_2]\cup[x'_3, x'_4]$ at $u=u_c$: schematic pictures of minimal surfaces (red curves).  }\label{hee}
\end{figure}

From the holographic perspective, the entanglement entropy is identified with the minimal area of the surface anchored at the boundary points of the interval at the cutoff $u=u_c$.  For two intervals, there are two configurations of minimal surfaces: disconnected phase ($s$-channel) and connected phase ($t$-channel) as shown in Figure \ref{hee}.  By comparing two areas of the minimal surfaces, we can determine `phase transition' points of the holographic entanglement entropy. 
This phase structure of the holographic entanglement entropy will be useful to understand the entanglement entropy in the $T\overline{T}$ deformed CFT. 

In \cite{Ota:2019yfe}, the phase transition of the holographic entanglement entropy of two intervals with the {\it same} lengths  at {\it zero} temperature was investigated. In this paper, we generalize it to two intervals with {\it different} lengths and at {\it finite} temperature.  
At high temperature, our holographic computation shows that s-channel is always favored so there is no phase transition.
We show that this agrees with the field theory computations. 
However, at zero temperature and intermediate temperature there is a phase transition between $s$-channel and $t$-channel.  We also discuss the cutoff dependence of  the phase transition of the entanglement entropy.

The organization of this paper is as follows. In section \ref{section2}, we provide the formulas of the R\'{e}nyi entropy with the first order $T\overline{T}$ deformation by using the twist operators. Based on this formulas, in section \ref{section3} and  \ref{section4},  we explicitly compute the entanglement entropy  and the R\'{e}nyi entropy respectively.  In section \ref{section5},  we study the holographic entanglement entropy of two intervals with the finite radius cutoff and its phase structure.  We conclude in section \ref{summary}.

\section{Formulas of the R\'{e}nyi  entropy in the deformed CFT }\label{section2}
In this section, we develop a formalism to compute the R\'{e}nyi  entropy in 2d CFT at finite temperature with a first order perturbation by the $T\overline{T}$ deformation. We use the twist operators to compute correlation functions on the $n$-sheeted surface in the replica method. This formalism can generalize the computation of the R\'{e}nyi entropy of a single interval \cite{Chen:2018eqk} to multiple intervals.

\begin{figure}[]
\centering
     {\includegraphics[width=10cm]{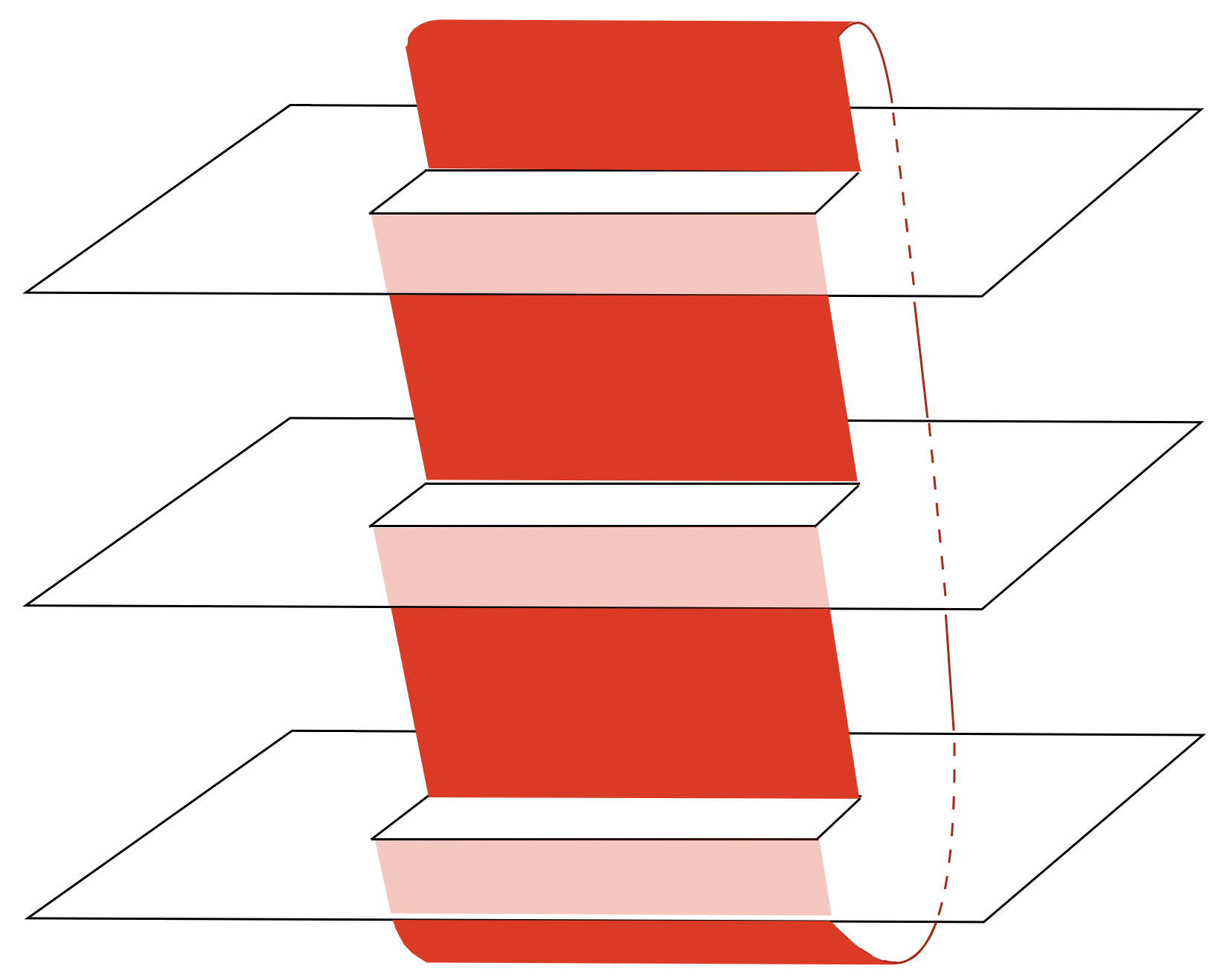} \label{}}
 \caption{Schematic picture of manifold $\mathcal{M}^{3}$.}\label{ManifoldFig}
\end{figure}

Consider the deformed CFT by $T\overline{T}$ deformation living on the manifold $\mathcal{M}$.
By the replica method \cite{Calabrese:2004eu, Calabrese:2009qy}, the R\'{e}nyi entropy of a subsystem $A\in\mathcal{M}$ can be expressed as follows
\begin{align}
S_n(A):=\frac{1}{1-n}\log\frac{Z_n(A)}{Z^n},\label{ee1}
\end{align}
where $Z$ is the partition function defined on $\mathcal{M}$ and $Z_n(A)$ is the one defined on the $n$-sheeted surface $\mathcal{M}^n(A)$ which is constructed from sewing $n$ copies of $\mathcal{M}$ cyclically along $A$ on each $\mathcal{M}$.
A vivid example of this n-sheeted surface is displayed in Fig. \ref{ManifoldFig}.
Note that this R\'{e}nyi entropy \eqref{ee1} reduces to the entanglement entropy in the $n\rightarrow1$ limit:
\begin{align}
S(A)=\lim_{n\to1}S_n(A)\,.
\end{align}

In this paper, we consider the deformed CFT by the first order perturbation of $T\overline{T}$ {at finite temperature, i.e. on a cylinder $\mathcal{M}$.}  The perturbative action is
\begin{align}
S_\textrm{QFT}=S_{\textrm{CFT}}+\mu\int_\mathcal{M}T\overline{T},\label{Sqft}
\end{align}
where $T$ and $\overline{T}$ are the energy momentum tensors in the undeformed CFT. We use coordinate $w=x+i\tau$ and $\bar{w}=x-i\tau$ on the cylinder $\mathcal{M}$. Here, $\tau$ is periodic as $\tau\sim\tau+\beta$, and $\beta$ can be interpreted as the inverse temperature. Thus, the integral sign in (\ref{Sqft}) is understood as $\int_{\mathcal{M}}:=\int^{\infty}_{-\infty} \dd x\int^{\beta}_0 \dd\tau$.      With this perturbative action, the first order perturbation of $S_n(A)$ is \cite{Chen:2018eqk}
\begin{align}
\delta S_n(A)=\frac{\mu}{n-1}\left(\int_{\mathcal{M}^n}\langle T\overline{T}\rangle_{\mathcal{M}^n}-n\int_{\mathcal{M}} \langle T\overline{T}\rangle_{\mathcal{M}}\right).\label{ee2}
\end{align}
Since we consider the first order perturbation by $\mu$, we can use CFT techniques to compute the correlation functions in (\ref{ee2}). 

Let us express $\int_{\mathcal{M}^n}\langle T\overline{T}\rangle_{\mathcal{M}^n}$ by the twist operators. Consider $m$ intervals $A=[x_1, x_2]\cup\cdots\cup[x_{2m-1}, x_{2m}]$  as the subsystem. In this subsystem, $\int_{\mathcal{M}^n}\langle T\overline{T}\rangle_{\mathcal{M}^n}$ is given by \cite{Cardy:2007mb, Calabrese:2009qy, Sun:2019ijq}
\begin{align}
\begin{split}
\int_{\mathcal{M}^n}\langle T\overline{T}\rangle_{\mathcal{M}^n} 
&= \sum_{k=1}^{n} \int_\mathcal{M} \frac{\langle T_{k}(w)\overline{T}_{k}(\bar{w})\prod_{i=1}^m \sigma_n(w_{2i-1}, \bar{w}_{2i-1})\bar{\sigma}_n(w_{2i}, \bar{w}_{2i})\rangle_{\mathcal{M}}}{\langle\prod_{i=1}^m\sigma_n(w_{2i-1}, \bar{w}_{2i-1})\bar{\sigma}_n(w_{2i}, \bar{w}_{2i})\rangle_{\mathcal{M}}} \\
&=\int_\mathcal{M}\frac{1}{n}\frac{\langle T^{(n)}(w)\overline{T}^{(n)}(\bar{w})\prod_{i=1}^m\sigma_n(w_{2i-1}, \bar{w}_{2i-1})\bar{\sigma}_n(w_{2i}, \bar{w}_{2i})\rangle_{\mathcal{M}}}{\langle\prod_{i=1}^m\sigma_n(w_{2i-1}, \bar{w}_{2i-1})\bar{\sigma}_n(w_{2i}, \bar{w}_{2i})\rangle_{\mathcal{M}}} \,,  \label{ee3}
\end{split}
\end{align}
where $\sigma_n$ and $\bar{\sigma}_n$ are the twist operators and $w_{i}$ denotes an end point of each interval.
Note that $k$ in the first line is a replica index and $T_{k}(w)\overline{T}_{k}(\bar{w})$ is defined from the $k$-th replica fields. In the second line, we  use the following identity in a correlation function: 
\begin{align}
\begin{split}
\sum_{k=1}^{n} \, \langle T_{k}(w) \, \overline{T}_{k}(\bar{w}) \,  \cdots   \rangle =  \frac{1}{n} \, \langle T^{(n)}(w) \, \overline{T}^{(n)}(\bar{\omega}) \,  \cdots  \rangle \,,
\label{REPLICA}
\end{split}
\end{align}
where $T^{(n)}(w)$ and $\overline{T}^{(n)}(\bar{w})$ are the total energy momentum tensors of $n$ replica fields, which are defined as follows:
\begin{align}
\begin{split}
T^{(n)}(w)  := \sum_{k=1}^{n} \, T_{k}(w) \,, \qquad  \overline{T}^{(n)}(\bar{w})  := \sum_{k=1}^{n} \, \overline{T}_{k}(\bar{w}) \,.
\label{REPLICA2}
\end{split}
\end{align}
Note the \eqref{REPLICA} is valid when the operators ``$\cdots$''  therein have the cyclic symmetry under the change of replica indices.

In order to compute the correlation functions on the cylinder $\mathcal{M}$, consider a conformal map
\begin{align}
z=e^{\frac{2\pi w}{\beta}}\,, \qquad \bar{z}=e^{\frac{2\pi \bar{w}}{\beta}},
\end{align}
from $w$ on $\mathcal{M}$ to $z$ on a complex plane $\mathcal{C}$. Under this transformation, the total energy momentum tensors of $n$ replica fields transform as
\begin{equation}
\begin{split}
&T^{(n)}(w)=\left(\frac{2\pi}{\beta}z\right)^2T^{(n)}(z)-\frac{\pi^2nc}{6\beta^2}, \\
&\overline{T}^{(n)}(\bar{w})=\left(\frac{2\pi}{\beta} \bar{z}\right)^2\overline{T}^{(n)}(\bar{z})-\frac{\pi^2nc}{6\beta^2},\label{temt}
\end{split}
\end{equation}
{where $c$ is the central charge of the undeformed CFT. }
For the calculation of \eqref{ee3}, we use (\ref{temt}) and the Ward identity with the energy momentum tensor (see, for example, \cite{Guica:2019vnb}) 
\begin{align}
\begin{split}
\langle T^{(n)}(z)\mathcal{O}_1(z_1, \bar{z}_1) \cdots & \mathcal{O}_{2m}(z_{2m}, \bar{z}_{2m}) \rangle_\mathcal{C} \\
& =\sum_{j=1}^{2m}\left(\frac{h_j}{(z-z_j)^2}+\frac{1}{z-z_j}\partial_{z_j}\right)\langle \mathcal{O}_1(z_1, \bar{z}_1)\cdots\mathcal{O}_{2m}(z_{2m}, \bar{z}_{2m})\rangle_\mathcal{C}, \\
\langle \overline{T}^{(n)}(\bar{z})\mathcal{O}_1(z_1, \bar{z}_1) \cdots & \mathcal{O}_{2m}(z_{2m}, \bar{z}_{2m})\rangle_\mathcal{C} \\
& = \sum_{j=1}^{2m}\left(\frac{\bar{h}_j}{(\bar{z}-\bar{z}_j)^2}+\frac{1}{\bar{z}-\bar{z}_j}\partial_{\bar{z}_j}\right)\langle \mathcal{O}_1(z_1, \bar{z}_1)\cdots\mathcal{O}_{2m}(z_{2m}, \bar{z}_{2m})\rangle_\mathcal{C},
\end{split}
\end{align}
where $h_i$ and $\bar{h}_i$ are the conformal dimensions of primary operators $\mathcal{O}_i$\footnote{In the usual conformal ward identity, the number of operators $\mathcal{O}_i$ could be any number. But here we restrict it as an even number $2m$ because our interest in this paper is the $m$-intervals.}.
The conformal dimensions of the twist operators are \cite{Calabrese:2004eu, Calabrese:2009qy}
\begin{align} \label{CDC}
h_{\sigma_n} =\bar{h}_{\sigma_n}=\frac{c}{24}\left(n-\frac{1}{n}\right).
\end{align}
Then, we obtain
\begin{align}
&\frac{\langle T^{(n)}(w)\overline{T}^{(n)}(\bar{w})\prod_{i=1}^m\sigma_n(w_{2i-1}, \bar{w}_{2i-1})\bar{\sigma}_n(w_{2i}, \bar{w}_{2i})\rangle_{\mathcal{M}}}{\langle\prod_{i=1}^m\sigma_n(w_{2i-1}, \bar{w}_{2i-1})\bar{\sigma}_n(w_{2i}, \bar{w}_{2i})\rangle_{\mathcal{M}}}\notag\\
=&\frac{1}{\langle\prod_{i=1}^m\sigma_n(z_{2i-1}, \bar{z}_{2i-1})\bar{\sigma}_n(z_{2i}, \bar{z}_{2i})\rangle_{\mathcal{C}}}\left[-\frac{\pi^2nc}{6\beta^2}+\left(\frac{2\pi}{\beta}z\right)^2\sum_{j=1}^{2m}\left(\frac{c(n-\frac{1}{n})}{24(z-z_j)^2}+\frac{1}{z-z_j}\partial_{z_j}\right)\right]\notag\\
\times&\left[-\frac{\pi^2nc}{6\beta^2}+\left(\frac{2\pi}{\beta}\bar{z}\right)^2\sum_{j=1}^{2m}\left(\frac{c(n-\frac{1}{n})}{24(\bar{z}-\bar{z}_j)^2}+\frac{1}{\bar{z}-\bar{z}_j}\partial_{\bar{z}_j}\right)\right]
\langle\prod_{i=1}^m\sigma_n(z_{2i-1}, \bar{z}_{2i-1})\bar{\sigma}_n(z_{2i}, \bar{z}_{2i})\rangle_{\mathcal{C}},
\label{ee5}
\end{align}
where the differential operators $\partial_{z_j}$ and $\partial_{\bar{z}_j}$ in (\ref{ee5}) act only on the correlation function. 

Finally, with (\ref{ee2}), (\ref{ee3}), and (\ref{ee5}), we obtain an expression of the R\'{e}nyi  entropy in the deformed CFT of the multiple intervals by using the twist operators
\begin{align}
&\delta S_n(A)\notag = - \frac{\mu c}{12(n-1)}\frac{8\pi^4}{\beta^4} \\
&\times \int_{\mathcal{M}}\Biggm[z^2\sum_{j=1}^{2m}\left(\frac{c(n-1/n)}{24(z-z_j)^2}+\frac{\partial_{z_j}\log\langle\prod_{i=1}^m\sigma_n(z_{2i-1}, \bar{z}_{2i-1})\bar{\sigma}_n(z_{2i}, \bar{z}_{2i})\rangle_{\mathcal{C}}}{(z-z_j)}\right)\notag\\
&\;\;\;\;\;\;\;\;\;\;+\bar{z}^2\sum_{j=1}^{2m}\left(\frac{c(n-1/n)}{24(\bar{z}-\bar{z}_j)^2}+\frac{\partial_{\bar{z}_j}\log\langle\prod_{i=1}^m\sigma_n(z_{2i-1}, \bar{z}_{2i-1})\bar{\sigma}_n(z_{2i}, \bar{z}_{2i})\rangle_{\mathcal{C}}}{(\bar{z}-\bar{z}_j)}\right)\Biggm]\notag\\
&+\frac{\mu}{n(n-1)}\frac{16\pi^4}{\beta^4}\frac{1}{\langle\prod_{i=1}^m\sigma_n(z_{2i-1}, \bar{z}_{2i-1})\bar{\sigma}_n(z_{2i}, \bar{z}_{2i})\rangle_{\mathcal{C}}}\notag\\
&\times\int_{\mathcal{M}}z^2\left(\sum_{j=1}^{2m}\left(\frac{c(n-1/n)}{24(z-z_j)^2}+\frac{\partial_{z_j}}{(z-z_j)}\right)\right)
\bar{z}^2\left(\sum_{j=1}^{2m}\left(\frac{c(n-1/n)}{24(\bar{z}-\bar{z}_j)^2}+\frac{\partial_{\bar{z}_j}}{(\bar{z}-\bar{z}_j)}\right)\right)\notag\\
&\times\langle\prod_{i=1}^m\sigma_n(z_{2i-1}, \bar{z}_{2i-1})\bar{\sigma}_n(z_{2i}, \bar{z}_{2i})\rangle_{\mathcal{C}}.\label{dsna2}
\end{align}
Taking the limit $n\to1$, the entanglement entropy $\delta S(A):=\lim_{n\to1}\delta S_n(A)$ is given by 
\begin{align}
&\delta S(A)\notag = -\mu \left(\frac{c}{12}\right)^2\frac{8\pi^4}{\beta^4} \\
&\;\;\; \times \int_{\mathcal{M}}\Biggm[z^2\sum_{j=1}^{2m}\left(\frac{1}{(z-z_j)^2}+\lim_{n\to1}\frac{12\partial_{z_j}\log\langle\prod_{i=1}^m\sigma_n(z_{2i-1}, \bar{z}_{2i-1})\bar{\sigma}_n(z_{2i}, \bar{z}_{2i})\rangle_{\mathcal{C}}}{c(n-1)(z-z_j)}\right)\notag\\
&\;\;\;\;\;\;\;\;\;\;\;\;\;+\bar{z}^2\sum_{j=1}^{2m}\left(\frac{1}{(\bar{z}-\bar{z}_j)^2}+\lim_{n\to1}\frac{12\partial_{\bar{z}_j}\log\langle\prod_{i=1}^m\sigma_n(z_{2i-1}, \bar{z}_{2i-1})\bar{\sigma}_n(z_{2i}, \bar{z}_{2i})\rangle_{\mathcal{C}}}{c(n-1)(\bar{z}-\bar{z}_j)}\right)\Biggm].\label{dsa}
\end{align}
%

If the correlation function is factorized as 
\begin{align}
\langle\prod_{i=1}^m\sigma_n(z_{2i-1}, \bar{z}_{2i-1})\bar{\sigma}_n(z_{2i}, \bar{z}_{2i})\rangle_{\mathcal{C}}=f(z_1, \cdots, z_{2m})g(\bar{z}_1, \cdots, \bar{z}_{2m}),\label{fcf}
\end{align}
we can obtain a simpler expression of $\delta S_n(A)$
\begin{align}
&\delta S_n(A)\notag\\
=&- \frac{\mu c}{12(n-1)}\frac{8\pi^4}{\beta^4}\int_{\mathcal{M}}\Biggm[z^2\sum_{j=1}^{2m}\left(\frac{c(n-1/n)}{24(z-z_j)^2}+\frac{\partial_{z_j}\log\langle\prod_{i=1}^m\sigma_n(z_{2i-1}, \bar{z}_{2i-1})\bar{\sigma}_n(z_{2i}, \bar{z}_{2i})\rangle_{\mathcal{C}}}{(z-z_j)}\right)\notag\\
&\;\;\;\;\;\;\;\;\;\;\;\;\;\;\;\;\;\;\;\;\;\;\;\;\;\;\;\;\;+\bar{z}^2\sum_{j=1}^{2m}\left(\frac{c(n-1/n)}{24(\bar{z}-\bar{z}_j)^2}+\frac{\partial_{\bar{z}_j}\log\langle\prod_{i=1}^m\sigma_n(z_{2i-1}, \bar{z}_{2i-1})\bar{\sigma}_n(z_{2i}, \bar{z}_{2i})\rangle_{\mathcal{C}}}{(\bar{z}-\bar{z}_j)}\right)\Biggm]\notag\\
&+\frac{\mu}{n(n-1)}\frac{16\pi^4}{\beta^4}\int_{\mathcal{M}}z^2\sum_{j=1}^{2m}\left(\frac{c(n-1/n)}{24(z-z_j)^2}+\frac{\partial_{z_j}\log\langle\prod_{i=1}^m\sigma_n(z_{2i-1}, \bar{z}_{2i-1})\bar{\sigma}_n(z_{2i}, \bar{z}_{2i})\rangle_{\mathcal{C}}}{(z-z_j)}\right)\notag\\
&\;\;\;\;\;\;\;\;\;\;\;\;\;\;\;\;\;\;\;\;\;\;\;\;\;\;\;\;\times \bar{z}^2\sum_{j=1}^{2m}\left(\frac{c(n-1/n)}{24(\bar{z}-\bar{z}_j)^2}+\frac{\partial_{\bar{z}_j}\log\langle\prod_{i=1}^m\sigma_n(z_{2i-1}, \bar{z}_{2i-1})\bar{\sigma}_n(z_{2i}, \bar{z}_{2i})\rangle_{\mathcal{C}}}{(\bar{z}-\bar{z}_j)}\right).\label{dsna}
\end{align}
The correlation functions which are studied explicitly in section \ref{section3} and \ref{section4} satisfy (\ref{fcf}). 
{From here, we set all $\tau_i$ are the same, which is equivalent to set $\tau_i =0$ because of the periodicity of $\tau$. }

\section{Explicit computation of the entanglement entropy $\delta S(A)$ }\label{section3}
In this section, we explicitly estimate $\delta S(A)$ (\ref{dsa}) of a single interval, two intervals, and multiple intervals. We show that $\delta S(A)$ of multiple intervals is a summation of $\delta S(A)$ of the single interval if the correlation function of multi intervals of the twist operators is factorized into the two point functions, such as a dominant contribution from the vacuum conformal block in the holographic CFT. This property of $\delta S(A)$ is consistent with the holographic entanglement entropy. 

\subsection{Single interval}
Let us first compute $\delta S(A)$ of a single interval case by the twist operator method (\ref{dsa}). 
 The correlation function of the twist operators for the single interval $A=[x_1, x_2]$ is
\begin{align}
\langle\sigma_n(z_1, \bar{z}_1)\bar{\sigma}_n(z_2, \bar{z}_2)\rangle_{\mathcal{C}} &= \frac{c_n}{|z_1-z_2|^{2(h_{\sigma_{n}}+\bar{h}_{\sigma_{n}})}}
= \frac{c_{n}}{|z_1-z_2|^{\frac{c}{6}\left(n-\frac{1}{n}\right)}} \,, \label{tp}
\end{align}
where $c_n$ is a constant and \eqref{CDC} is used.
Thus, $\delta S(A)$ (\ref{dsa}) of the single interval is
\begin{align}
\begin{split}
\delta S(A) & \\
=&-\mu \left(\frac{c}{12}\right)^2 \frac{8\pi^4}{\beta^4} \int_{\mathcal{M}} \Biggl[z^2 \biggl(\frac{1}{(z-z_1)^2} + \frac{1}{(z-z_2)^2} \\
  &\qquad\qquad\qquad\qquad\qquad\quad +\frac{-2}{(z-z_1)(z_1-z_2)}+\frac{-2}{(z-z_2)(z_2-z_1)}\biggr) + \textrm{h.c.}\Biggr] \\
=&-\mu \left(\frac{c}{12}\right)^2\frac{8\pi^4}{\beta^4}\int_{\mathcal{M}}\left(\frac{z^2(z_1-z_2)^2}{(z-z_1)^2(z-z_2)^2}+\textrm{h.c.}\right) \\
=& - \mu \, \frac{\pi ^4 c^2 \, (x_{2}-x_{1})}{9 \beta ^3}\coth \left(\frac{\pi  (x_{2}-x_{1}) }{\beta }\right) \label{dsasi} \,,
\end{split}
\end{align}
where we used $z_{i} := e^{\frac{2\pi}{\beta}x_{i}}$ with $\tau_i=0$ in the last equality. For a detailed calculation of the integration in \eqref{dsasi}, see Appendix \ref{appendixA}.
Note that  $``-2"$ in the  numerator of the first equality comes from the derivative of the logarithm term in \eqref{dsa} i.e. we only need to read off the power of $z_1-z_2$ in \eqref{tp}, $\frac{-c}{12}(n-\frac{1}{n}).$\footnote{It is not $\frac{c}{6}(n-\frac{1}{n})$ because $|z_1-z_2| = \sqrt{(z_1-z_2)(\bar{z}_1-\bar{z}_2)}$.  }
Finally, by considering the remaining factor $\frac{12}{c(n-1)}$ in \eqref{dsa} we have $\frac{12}{c(n-1)}  \times  \frac{-c}{12}(n-\frac{1}{n}) = -\frac{n+1}{n}$ which becomes $``-2"$ in the $n\rightarrow1$ limit.
Taking $x_{1}=0$, $x_2=\ell$ in \eqref{dsasi}, we reproduce the same result in \cite{Chen:2018eqk}, where a different approach is used: a conformal map between $\mathcal{M}^n$ and $\mathcal{C}$.

\subsection{Two intervals}\label{sec32}
Let us turn to compute $\delta S(A)$ of two intervals $A=[x_1, x_2]\cup[x_3, x_4]$. The correlation function of the twist operators for  two intervals is the four point function $\langle\sigma_n(z_1, \bar{z}_1)\bar{\sigma}_n(z_2, \bar{z}_2)$ \\ $\sigma_n(z_3, \bar{z}_3)\bar{\sigma}_n(z_4, \bar{z}_4)\rangle_{\mathcal{C}}$. Generally, four point functions in CFT are not universal and depend on the details of CFT. Thus, to proceed, we consider the {\it holographic CFT}, because 
in the holographic CFT, it was argued that the vacuum conformal block is a dominant contribution in the four point function of the twist operators~\cite{Hartman:2013mia}\footnote{Rigorously speaking, there is a possibility that the vacuum conformal block is not dominant in some region of $\eta$ defined in  \eqref{crosr}. In this paper, we assume that the vacuum conformal block is dominant in the entire region $0\le\eta\le1$. }. Furthermore,  the four point function in the limit $n\to1$ is factorized into the two point functions in the leading order (see, also \cite{Fitzpatrick:2014vua, Perlmutter:2015iya}).  

{In more detail, let us consider the four point function in the holographic CFT classified by the cross ratio\footnote{In this paper, $\eta$ is real such that $\eta=\bar{\eta}$.} $\eta$
\begin{equation} \label{crosr}
\eta := \frac{(z_1-z_2)(z_3-z_4)}{(z_1-z_3)(z_2-z_4)} \,. 
\end{equation}
This cross ratio is invariant under the global conformal transformation~\cite{DiFrancesco:1997nk} and it plays a role in classifying convergence of the conformal block expansion in each channel~(see, for example, \cite{Hartman:2015lfa}). We will call the region of $0\le\eta\le1/2$ ``$s$-channel" and the region of $1/2\le\eta\le1$ ``$t$-channel".  

The four point function can be approximated by the vacuum conformal block as\footnote{We omit  constant terms in the normalization of the correlation functions.
 }, for example in the $s$-channel\footnote{For the $t$-channel, we exchange $z_2 \leftrightarrow z_4$ (in this case $\eta$ becomes $1-\eta).$} 
 \cite{Fitzpatrick:2014vua, Perlmutter:2015iya}
\begin{align}
\begin{split}
\log\langle\sigma_n(z_{1}, \bar{z}_{1})\bar{\sigma}_n(z_{2}, \bar{z}_{2})\sigma_n(z_{3}, \bar{z}_{3})\bar{\sigma}_n(z_{4}, \bar{z}_{4})\rangle_{\mathcal{C}} \,\sim\, -\frac{nc}{6}f_\text{vac}(z_1,z_2,z_3,z_4) \,+\, \text{h.c} \,,
\end{split}
\end{align}
with
\begin{align}
\begin{split}
f_\text{vac}(z_1,z_2,z_3,z_4) &=  \epsilon_n\left(2\log[z_2-z_1] + 2\log[z_4-z_3]-\frac{\epsilon_n}{3}\eta^2\;_2F_1(2, 2; 4; \eta)\right) \\
& \quad + \mathcal{O}\left(\left(n-1\right)^3\right), \\
\epsilon_n &:=  \frac{6}{nc}h_{\sigma_n} = \frac{n+1}{4n^2}\left(n-1\right)\,, \label{fp}
\end{split}
\end{align}
where $ _2F_1(\alpha, \beta; \gamma; \delta)$ is the hypergeometric function and $h_{\sigma_n}$ is defined in \eqref{CDC}. }
 In the limit $n\to1$, we can ignore $\epsilon_n^2 \, \eta^2 \, _2F_1(2, 2; 4; \eta)$ and the higher order terms of $n-1$ in (\ref{fp}). Thus, for the case $n\to1$, the four point function in $s$-channel $(0\le\eta\le1/2)$ is  factorized as
\begin{align}
\langle\sigma_n(z_1, \bar{z}_1)\bar{\sigma}_n(z_2, \bar{z}_2)\sigma_n(z_3, \bar{z}_3)\bar{\sigma}_n(z_4, \bar{z}_4)\rangle_{\mathcal{C}}\sim\frac{1}{|z_1-z_2|^{2(h_{\sigma_{n}}+\bar{h}_{\sigma_{n}})}|z_3-z_4|^{2(h_{\sigma_{n}}+\bar{h}_{\sigma_{n}})}}. \label{sc}
\end{align}
Similarly, the four point function in $t$-channel $(1/2\le\eta\le1)$ is  factorized as
\begin{align}
\langle\sigma_n(z_1, \bar{z}_1)\bar{\sigma}_n(z_2, \bar{z}_2)\sigma_n(z_3, \bar{z}_3)\bar{\sigma}_n(z_4, \bar{z}_4)\rangle_{\mathcal{C}}\sim\frac{1}{|z_1-z_4|^{2(h_{\sigma_{n}}+\bar{h}_{\sigma_{n}})}|z_3-z_2|^{2(h_{\sigma_{n}}+\bar{h}_{\sigma_{n}})}}. \label{tc}
\end{align}

Substituting (\ref{sc}) and  (\ref{tc}) into (\ref{dsa}), we obtain
\begin{align}
\begin{split}
\delta S_{\,\text{s-ch}}(A) & \;\; \\
\sim&-\mu \left(\frac{c}{12}\right)^2\frac{8\pi^4}{\beta^4}\int_{\mathcal{M}} \Biggl[ \left(\frac{z^2(z_1-z_2)^2}{(z-z_1)^2(z-z_2)^2}+\textrm{h.c.}\right) + \left(\frac{z^2(z_3-z_4)^2}{(z-z_3)^2(z-z_4)^2}+\textrm{h.c.}\right) \Biggr] \\
=& - \mu \, \frac{\pi ^4 c^2 \, (x_{2}-x_{1})}{9 \beta ^3}\coth \left(\frac{\pi  (x_{2}-x_{1}) }{\beta }\right)  - \mu \, \frac{\pi ^4 c^2 \, (x_{4}-x_{3})}{9 \beta ^3}\coth \left(\frac{\pi  (x_{4}-x_{3}) }{\beta }\right) \label{scds}
\end{split}
\end{align}
and
\begin{equation}
\delta S_{\,\text{t-ch}}(A) =  \quad   z_2 \leftrightarrow  z_4  \quad  \mathrm{and}  \quad x_2 \leftrightarrow  x_4 \quad \mathrm{in} \quad \delta S_{\,\text{s-ch}}(A) \,.  \label{tcds}
\end{equation}
Eqs. \eqref{scds} and \eqref{tcds} can be considered as a double summation of \eqref{dsasi}.
Therefore, in the deformed holographic CFT, $\delta S(A)$ of two intervals is the summation over $\delta S(A)$ of the single interval.
Indeed, this additive property comes from the $\log$ term in \eqref{dsa} with the factorization in \eqref{sc} and \eqref{tc}.

{The author} in~\cite{Hartman:2013mia} {argued} that, {by using the vacuum conformal block approximation,} $\delta S(A)$ of the two intervals in the \textit{undeformed} {holographic} CFT is the summation of the single interval case. This additive property is also shown in~\cite{Ryu:2006bv, Ryu:2006ef, Headrick:2010zt} within a holographic framework without introducing cutoff.
Here, we have shown that this additive property still holds in the $T\overline{T}$ deformed {holographic} CFT by the field theory computation from \eqref{scds} and \eqref{tcds}. In section \ref{section5}, we will show this additive property works also in the holographic theory (with a finite cutoff dual to $T\overline{T}$ deformation) and is consistent with the field theory results here.  

%


\subsection{Multiple intervals}
Finally, we compute $\delta S(A)$ of multiple intervals $A=[x_1, x_2]\cup\cdots\cup[x_{2m-1}, x_{2m}]$. Consider the holographic CFT in which the correlation function of the twist operators in the limit $n\to1$ is factorized into the two point functions because of the dominant vacuum conformal block \cite{Hartman:2013mia} as
\begin{align}
\langle\prod_{i=1}^m\sigma_n(z_{2i-1}, \bar{z}_{2i-1})\bar{\sigma}_n(z_{2i}, \bar{z}_{2i})\rangle_{\mathcal{C}}\sim\prod_{i=1}^m\frac{1}{|z_{k_i}-z_{l_i}|^{2(h_{\sigma_{n}}+\bar{h}_{\sigma_{n}})}} \;\;\;\;(n\to1),\label{mi}
\end{align}
where $k_i$ and $l_i$ are determined by configuration of the multiple intervals in the same manner as the correlation function for the two intervals. Substituting (\ref{mi}) into (\ref{dsa}), we obtain
\begin{align}
\begin{split}
\delta S(A)\sim&-\sum_{i=1}^m\mu \left(\frac{c}{12}\right)^2\frac{8\pi^4}{\beta^4}\int_{\mathcal{M}}\left(\frac{z^2(z_{k_i}-z_{l_i})^2}{(z-z_{k_i})^2(z-z_{l_i})^2}+\textrm{h.c.}\right)\\
=& - \sum_{i=1}^m \mu \, \frac{\pi ^4 c^2 \, (x_{k_{i}}-x_{\ell_{i}})}{9 \beta ^3}\coth \left(\frac{\pi  (x_{k_{i}}-x_{\ell_{i}}) }{\beta }\right).   \label{mids}
\end{split}
\end{align}
Therefore, $\delta S(A)$ of multiple intervals in the deformed holographic CFT which has the property (\ref{mi}) is the summation of $\delta S(A)$ of the single interval. This additive property comes from the $\log$ term in \eqref{dsa} with the factorization in \eqref{mi}. 
This property of $\delta S(A)$ is consistent with the Ryu-Takayanagi formula with the radius cutoff.


\section{Explicit computation of the R\'{e}nyi entropy $\delta S_n(A)$} \label{section4}
In this section, we evaluate the R\'{e}nyi entropy $\delta S_n(A)$. 
First, we compute $\delta S_n(A)$ of single interval case by using our twist operator method \eqref{dsna}.  Our result agrees with the one in~\cite{Chen:2018eqk}, where a different method was used. 
Second, we consider the two interval case and show that $\delta S_n(A)$ of two intervals can be expressed as a summation of a single interval case in some limit. Finally, we make some comments about a holographic interpretation of our results.

\subsection{Single interval}
Consider $\delta S_n(A)$ of a single interval $A=[x_1, x_2]$. Substituting (\ref{tp}) into (\ref{dsna}), we obtain 
\begin{align}
\begin{split}
\delta S_n(A) &\\
=&- \frac{(n+1)\mu }{2n}\left(\frac{c}{12}\right)^2\frac{8\pi^4}{\beta^4}\int_{\mathcal{M}}\left(\frac{z^2(z_1-z_2)^2}{(z-z_1)^2(z-z_2)^2}+\textrm{h.c.}\right)\\
&+\frac{(n+1)^2(n-1)\mu}{2n^3}\left(\frac{c}{12}\right)^2\frac{8\pi^4}{\beta^4}\int_{\mathcal{M}}\frac{z^2(z_1-z_2)^2}{(z-z_1)^2(z-z_2)^2}\frac{\bar{z}^2(\bar{z}_1-\bar{z}_2)^2}{(\bar{z}-\bar{z}_1)^2(\bar{z}-\bar{z}_2)^2}.
 \label{dsnasi}
\end{split}
\end{align}
{Here the first term comes from the first and second terms in \eqref{dsna} and the second term comes from the third term in \eqref{dsna}. This second term is a ``mixing term'' between holomorphic and anti-holomorphic part and will play an important role when we discuss the additivity of  $\delta S_n(A)$ for two intervals. 
Note also that it gives the same result as \eqref{dsasi} for $n=1$ as expected and in this case the mixing term vanishes. }

The integration in the second term in \eqref{dsnasi} needs a regularization procedure\footnote{See \cite{Chen:2018eqk} for the result after performing a regularization. In this paper, we will not do the regularization because our interest is to study the additive property of the R\'{e}nyi entropy.}. One can  see this divergence from the more general form in \eqref{RenyiResult}: By substituting $\bar{z}_{3} \rightarrow z_{1}$ and $\bar{z}_{4} \rightarrow z_{2}$ in \eqref{RenyiResult}, we find a divergency. The upshot of this twist field method is that our results for $\delta S_n(A)$ of a single interval \eqref{dsnasi} reproduces the same consequence reported in \cite{Chen:2018eqk}, {where a direct conformal map between $\mathcal{M}^n$ and $\mathcal{C}$ is used (while here we used a map between $\mathcal{M}$ and $\mathcal{C}$, and $\mathcal{M}^n$ and $\mathcal{M}$ are related by the twist operators.).}

\subsection{Two intervals}
Consider $\delta S_n(A)$ of two intervals $A=[x_1, x_2]\cup[x_3, x_4]$.  In general, $\delta S_n(A)$ of two intervals can not be expressed as a summation of a single interval case because of two reasons. 

{Firstly,  the four point function of the twist operators with $n\ne 1$ in the holographic CFT is not factorized into the two point functions as can be seen in \eqref{fp}, where $\epsilon_n \sim n-1$ but $n\ne1$. 
If we instead consider a limit that the cross ratio $\eta$ behaves as $\eta\to0$  or $\eta\to1$  the four point function is factorized into two point functions.  Note that this factorization is valid not only in holographic CFT but also in any CFT which obeys the cluster decomposition~\cite{Calabrese:2009ez, Headrick:2010zt}. 

Secondly, even if the four point function is factorized, $\delta S_n(A)$ of two intervals may not be the sum of the single interval's because of the ``mixing term'' between holomorphic and anti-holomorphic part, the third term in \eqref{dsna}. 
We will show that this mixing vanishes only for $\eta\to0$ but does not vanish for $\eta\to1$.
}

Let us first consider the limit $\eta\to0$. By substituting (\ref{sc}) into (\ref{dsna}), we obtain $\delta S_n(A)$ of the two interval:
%
\begin{equation}
\delta S_n(A)|_{\eta\to0}
\sim \,\, \delta S^{\,\,\,\text{single}}_n(A) \,+\, \delta S^{\,\,\,\text{single}}_n(A)\Bigr|_{z_{1} \,\rightarrow\, z_{3}, \,\, z_{2} \,\rightarrow\, z_{4}}  + \delta S^{\,\,\,\text{mixing}}_n(A) \,,
\label{dsnati}
\end{equation}
where $\delta S^{\,\,\,\text{single}}_n(A)$ denotes the R\'{e}nyi entropy of a single interval \eqref{dsnasi} and
\begin{equation} \label{mix987}
\begin{split}
 \delta S^{\,\,\,\text{mixing}}_n(A)  
 =&  \frac{(n+1)^2(n-1)\mu}{2n^3}\left(\frac{c}{12}\right)^2\frac{8\pi^4}{\beta^4} \\
 &\times \int_{\mathcal{M}}\left(\frac{z^2(z_1-z_2)^2}{(z-z_1)^2(z-z_2)^2}\frac{\bar{z}^2(\bar{z}_3-\bar{z}_4)^2}{(\bar{z}-\bar{z}_3)^2(\bar{z}-\bar{z}_4)^2}+\textrm{h.c.}\right) \,.
 \end{split}
\end{equation}

Note that if $n=1$, i.e. for entanglement entropy, the mixing term vanishes always. However, in general, if $n \ne 1$, it looks that $\delta S_n(A)$ of two intervals can not be expressed as a summation of a single interval case because of the mixing term \eqref{mix987}. 
To see when this mixing term is negligible, we compute the integral in Appendix \ref{appendixA} and it  boils down to 

\begin{equation} \label{mixing123}
\begin{split}
\frac{\delta S^{\,\,\,\text{mixing}}_n(A)}{c}   &= \left[\frac{(n+1)^2(n-1)}{n^3}   \frac{\pi^3}{72} \right] \, \,  \hat{\mu} \times \mathbb{M} \,, \\
&\hat{\mu} := \frac{\mu c}{\beta^2} \,, \\
&\mathbb{M} := \biggl( \frac{z_{1}^2 + \bar{z}_{3}^2}{(z_{1}-\bar{z}_{3})^2}   +  \frac{z_{1}^2+\bar{z}_{4}^2}{(z_{1}-\bar{z}_{4})^2}  +   \frac{z_{2}^2+\bar{z}_{3}^2}{(z_{2}-\bar{z}_{3})^2} +   \frac{z_{2}^2+\bar{z}_{4}^2}{(z_{2}-\bar{z}_{4})^2}  -    4     \\
&\qquad +  2\frac{(z_{1}+z_{2}) (\bar{z}_{3}+\bar{z}_{4})}{(z_{1}-z_{2}) (\bar{z}_{3}-\bar{z}_{4})}\ln \frac{(z_{1}-\bar{z}_{4}) (z_{2}-\bar{z}_{3})}{(z_{1}-\bar{z}_{3}) (z_{2}-\bar{z}_{4})}   \,  \biggr) \,,
\end{split}
\end{equation}
where $\hat{\mu}$ is a dimensionless parameter.
Let us choose the parameters  ($0 = x_1 < x_2 <x_3 < x_4 $) as 
\begin{equation}
 \ell_{12} := x_2 - x_1 \,, \quad  \ell_{23} := x_3 - x_2\,, \quad  \ell_{34} := x_4 - x_3 \,.
\end{equation}
If we fix $x_2$ and $\ell_{34}$; and take $x_3(x_4) \rightarrow \infty$,  it implies $z_1$ and $z_2$ are fixed and $\bar{z}_3(\bar{z}_4) \rightarrow \infty$ with the relation ($\tau_i=0$);
\begin{equation} \label{SubSti}
\begin{split}
\bar{z}_{4} = e^{\frac{2\pi}{\beta} x_{4}}
                   = e^{\frac{2\pi}{\beta} (x_{3} + \ell_{34})}
                   = \bar{z}_{3} \, e^{\frac{2\pi}{\beta} \ell_{34}} \,.
\end{split}
\end{equation}
Thus, the first five terms in $\mathbb{M}$ in \eqref{mixing123} sums up to zero. The last term also vanishes ($\sim \ln 1$) by itself.  
In other words, when two intervals are far from each other, $\delta S^{\,\,\,\text{mixing}}_n(A)$ vanishes, so $\delta S_n(A)$ of the two intervals is the summation of $\delta S_n(A)$ of the single interval.
{This is indeed the requirement $\eta \rightarrow 0$, which is the   condition we have already imposed to have a four point function factorized.}
On the other hand, if two intervals become close ($x_3 \rightarrow x_2$) the mixing term blows up because of the third term in $\mathbb{M}$.  

\begin{figure}[]
 \centering
      \subfigure[$\eta\to0$:  $\ell_{12} = 0.01, \, \beta = 2\pi$]
     {\includegraphics[width=7.3cm]{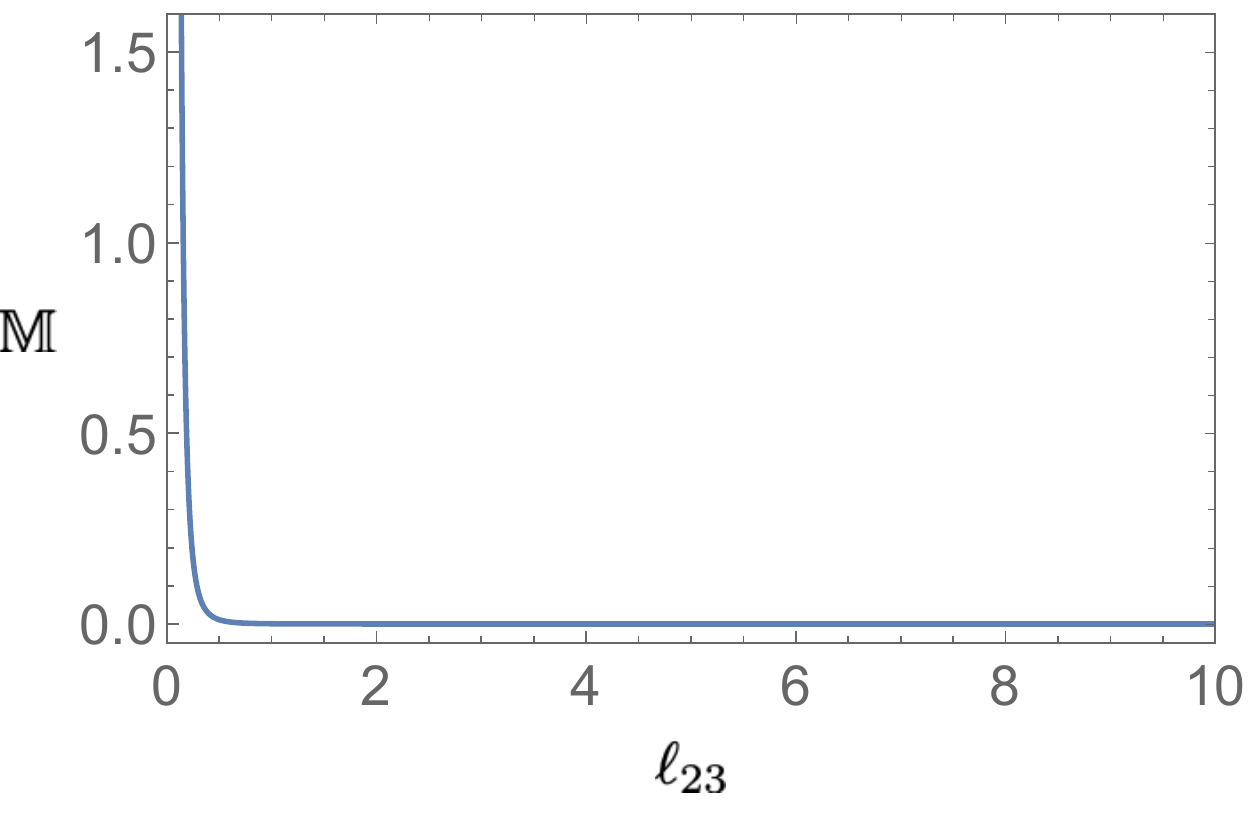} \label{CROSSFiga}}
      \subfigure[$\eta\to1$:  $\ell_{23} = 0.01, \, \beta = 2\pi$]
     {\includegraphics[width=7.3cm]{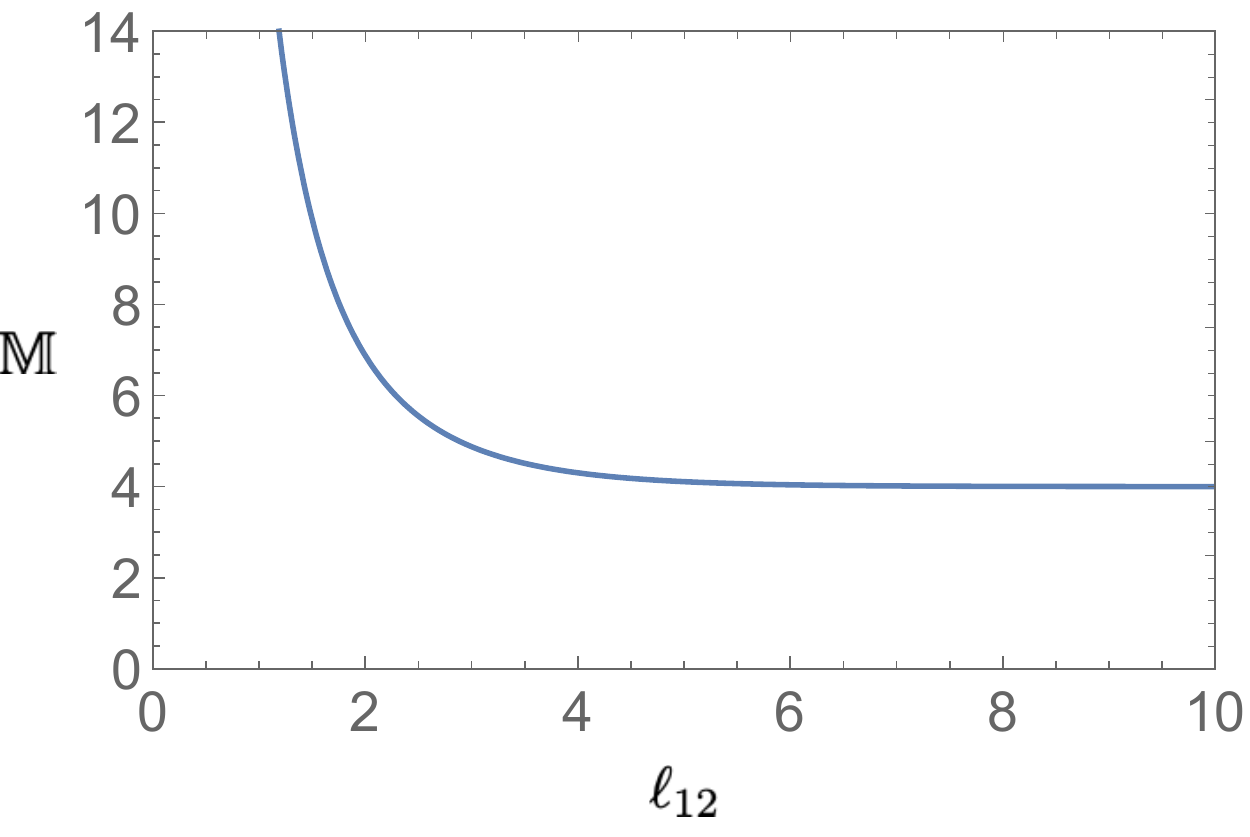} \label{CROSSFigb}}
          \caption{   {The mixing effect $\mathbb{M}$ in \eqref{mixing123} of symmetric configuration, $\ell_{12} = \ell_{34}$, where $\ell_{12} := |x_2 - x_1|$ and $\ell_{34} := |x_4 - x_3|$. } In order to satisfy the condition $\eta\to0$ (a) and $\eta\to1$ (b) only the ranges of $\ell_{23} \gg 1$ (a) or  $\ell_{23} \ll 1$ with a fixed $\ell_{12}$(b) are valid. } \label{CROSSFig}
\end{figure}

These two extreme limit will be interpolated as we dial $\ell_{23}$. To see this we make a plot of the $\mathbb{M}$ in \eqref{mixing123} in Fig. \ref{CROSSFiga}, where we choose 
\begin{equation}
{ \ell_{12}=  \ell_{34} = 0.01 \,, \quad \beta = 2\pi \,. }
\end{equation}
Because we are considering $\eta \rightarrow 0$ limit,  {$\ell_{23}/\ell_{12}$} should be large so only the range $\ell_{23} \gg 1$\footnote{More precisely, $\ell_{23}/(\beta/2\pi) \gg 1$. In fact, our choice of $\beta=2\pi$ corresponds to using the rescaled parameter $\tilde{\ell}_{ij} := 2\pi\ell_{ij}/\beta$. With this understanding, not to clutter, we use $\ell_{ij}$ without tilde.} is valid.

\begin{figure}[]
 \centering
      \subfigure[ $\eta\sim0$ ]
     {\includegraphics[width=7.3cm]{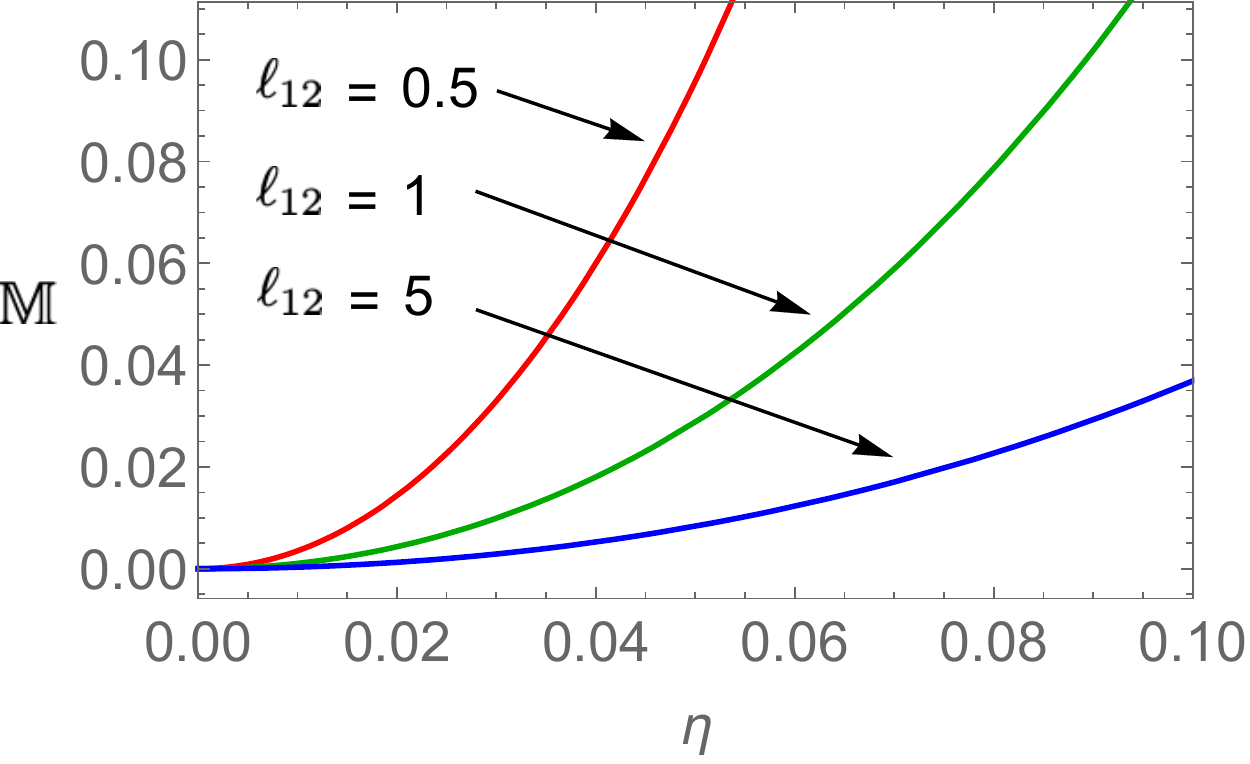} \label{}}
      \subfigure[ $\eta\sim1$ ]
     {\includegraphics[width=7.1cm]{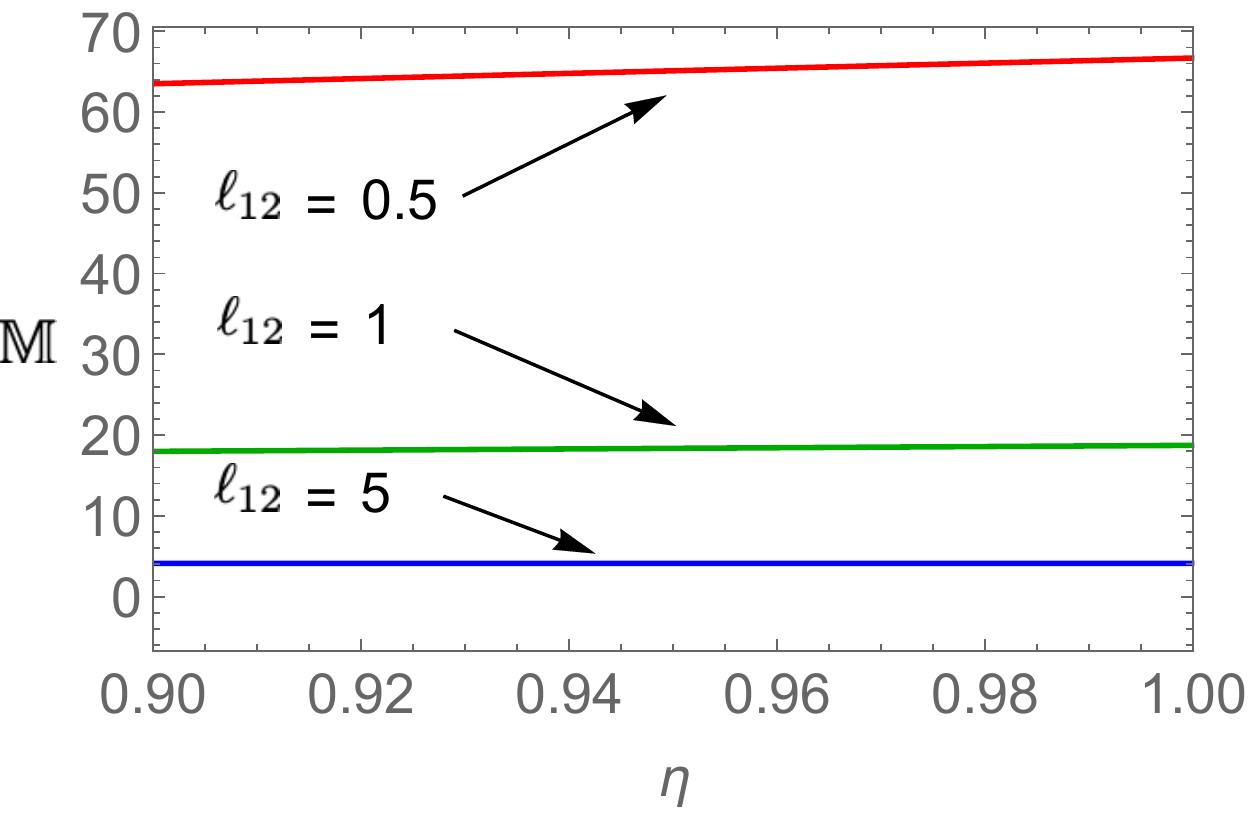} \label{}}
          \caption{The mixing effect $\mathbb{M}$ in \eqref{mixing123} vs $\eta$:  $\ell_{12}$ = 0.5, 1, 5 (red, green, blue).} \label{xxxc}
\end{figure}

{
In the limit $\eta\to1$, by using a relation of the correlation function in 2d CFT under $\eta\to 1-\eta$ as explained in  \cite{Calabrese:2009ez, Headrick:2010zt}, one can perform a similar analysis. Thus, $\delta S_n(A)$ of the two intervals is obtained by exchanging $z_{2} \leftrightarrow z_{4}$ in \eqref{dsnati}.
In Fig. \ref{CROSSFigb} we display the $\mathbb{M}$ in  \eqref{mixing123} after exchanging $z_{2} \leftrightarrow z_{4}$ with
\begin{equation}
\ell_{23} = 0.01 \,, \quad \beta = 2\pi \,.
\end{equation}
Here, we chose a small $\ell_{23}$ because we are considering $\eta \rightarrow 1$ limit, which is equivalent to {$\ell_{23} \ll 1$} with a fixed $\ell_{12}$.} 
$\mathbb{M}$ saturates to $4$ when $\ell_{12} = \ell_{34}$ increase while it  diverges if $\ell_{12} = \ell_{34}$ decreases.  This divergence is originated from the second and third term in $\mathbb{M}$ in \eqref{mixing123}, after exchanging $z_{2} \leftrightarrow z_{4}$.
Note that, roughly speaking, $\eta \to 1$ restricts the range of $\ell_{12} \gg 1 $.

{
We noted that, in Fig.~\ref{CROSSFig}, only the right part of both figures (large $\ell_{23}$ in (a) and large $\ell_{12}$ in (b)) is valid in order to satisfy the conditions $\eta \sim 0$ (a) and $\eta \sim 1$ (b). To demonstrate it more clearly we made a plot of the $\mathbb{M}$ in  \eqref{mixing123} versus $\eta$ in Fig.~\ref{xxxc} for $\ell_{12} = 0.5, 1, 5$.  If $\eta \sim 0$, $\mathbb{M}$ is very small and vanishes as $\eta \rightarrow 0$, while if $\eta \rightarrow 1$, $\mathbb{M}$ saturates to the finite value, which depends on $\ell_{12}$. The saturation value for a given $\ell_{12}$  is approximately the same as the value in Fig. \ref{CROSSFigb} since $\eta \rightarrow 1$ corresponds to $\ell_{23} \rightarrow 0$. There are two interesting facts in the limit of $\eta \rightarrow 1$:
\begin{itemize}
\item[(a)] $\mathbb{M}$ is non-zero
\item[(b)] $\mathbb{M}$ is saturated to the minimum value ($\sim 4$) as $\ell_{12}$ increases as shown in Fig. \ref{CROSSFigb}.
\end{itemize}
These two properties can be intuitively explained by holography. See the end of the next section.

However, there is one subtlety in our results\footnote{{We thank the referee for pointing out this issue.}}. The numerical value of $\mathbb{M}$ can be very large as $\ell_{12}$ becomes very small as shown in Fig.~\ref{xxxc}. In this case, the perturbation theory will break down if $\mathbb{M}$ reaches the value of order $1/\hat{\mu}\, $\footnote{{The numerical value of  the square bracket in the first line of \eqref{mixing123} is around $0.5$. It does not vary much as $n$ changes.}}. Note that this is problematic only if $\ell_{12}$ is small. Thus, we suspect that our perturbation theory may be justified by the existence of the lower bound of $\ell_{12}$ provided by the cut-off $\hat{\mu}$ scale. i.e. {because $\ell_{12}$ needs to be much longer than $\mu$}\footnote{{The effective field theory at small scale comparable to the deformation becomes non-local, where it is not clear that the entanglement entropy makes sense, e.g, see~\cite{Dubovsky:2012wk, Dubovsky:2013ira, Cooper:2013ffa}.}} it is bounded below so $\mathbb{M}$ may be well below $1/\hat{\mu}$. Currently, we do not have a more rigorous understanding on this issue and leave it as a future work.}

\subsection{Holographic interpretation}
The author of \cite{Dong:2016fnf} proposed the gravity dual of R\'{e}nyi  entropy in the holographic CFT as
\begin{align}
n^2\partial_n\left(\frac{n-1}{n}S_n(A)\right)=\frac{\textrm{Area}(\textrm{Cosmic Brane}_n)}{4G},
\end{align}
where $G$ is the Newton's constant, and the cosmic branes are anchored at the boundary of $A$.
In order to compute the area of the cosmic branes, we need to consider the back-reaction from the cosmic branes to the bulk geometry. Thus, generally, the holographic R\'{e}nyi  entropy of two intervals with $n\ne1$ is not summation of the one of the single interval because of the back-reaction between the two cosmic branes. 

However, if  two cosmic branes are far from each other, the back-reaction between them is negligible. {This is the limit of $\eta \rightarrow 0$.} In this limit, the holographic R\'{e}nyi  entropy of the two intervals becomes summation of the one of the single interval. Even though we introduce the radius cutoff, which corresponds to the $T\overline{T}$ deformation of the holographic CFT, this property of the holographic R\'{e}nyi  entropy will be still valid. Thus, it is consistent with our field theory result in the previous subsection.   

{Away from the limit $\eta \rightarrow 0$,}
$\delta S_n(A)$ of the two intervals in the deformed holographic CFT includes corrections to \eqref{mixing123} and higher order terms of $\eta$ in the vacuum conformal block. These corrections may be related to the back-reaction between the two cosmic branes in the holographic R\'{e}nyi entanglement entropy formula with the radius cutoff. 

{Let us turn to the holographic interpretation of the two properties (a) and (b) in the previous section.
For $\eta \rightarrow 1$, our field theory result shows, at given temperature $\beta$, there is a saturation of the mixing term when $\ell_{12}$ increases.  It may be understood holographically  as follows: i) the cosmic brane for the range $\ell_{23}$ is almost a point at the cutoff because $\eta \sim 1$ means $\ell_{23} \sim 0$ , ii) the cosmic brane for the range $\ell_{14}$ will be close to and bounded by the horizon if $\ell_{12} \gg 1 (\ell_{14} \sim 2\ell_{12})$. Thus, the effective distance between two cosmic branes is saturated, which is basically the distance between the cut-off and the horizon. Thus, the effect of the back-reaction is also saturated.  {Only if the cut-off or temperature becomes zero, the effective distance between two cosmic branes becomes infinite and their back-reaction may be negligible. This information is encoded in $\hat{\mu}$ in \eqref{mixing123}.} Note that if $\eta \rightarrow 0$, two cosmic branes are far away always so their interaction is negligible regardless of the cut-off.}

\section{Holographic entanglement entropy and phase transitions}\label{section5}

In this section, we study the holographic entanglement entropy with a finite cutoff and compare it with the previous field theory result. While the perturbative field theory is valid only for small deformations, the holographic method can be used for general deformations. 
We specify the parameter regime that the field theory and holographic results agree.  We also investigate the phase transitions between the $s$-channel and the $t$-channel for the two interval cases with a finite radius cutoff in the holographic framework.

\subsection{Holography: single interval}

Let us consider a planar  BTZ black hole:
\begin{align}
\dd s^2=\frac{r^2-r_h^2}{L^2}\dd t^2+\frac{L^2}{r^2-r^2_h}\dd r^2+\frac{r^2}{L^2} \dd \tilde{x}^2\,,
\end{align}
where $L$ is the AdS radius, and $r_h$ is the horizon radius.
At the cutoff radius $r=r_c$, 
\begin{equation}
\dd s^2 \sim \dd t^2+\frac{1}{1-r_h^2/r_c^2} \dd \tilde{x}^2\ =  \dd t^2+ \dd x^2 \,,
\end{equation}
where 
\begin{equation}
x := \frac{\tilde{x}}{\sqrt{1-r_h^2/r_c^2}} \,. 
\end{equation}

The holographic entanglement entropy of a single interval between $\tilde{x}=\tilde{x}_i$ and $\tilde{x}=\tilde{x}_j$ at the cutoff radius $r=r_c$ in this black hole geometry is~\cite{Ryu:2006bv, Chen:2018eqk}
\begin{equation} \label{PTEE}
\begin{split}
S^{H}(\ell_{ij}) &= \frac{L}{4G} \log\left( \mathcal{A}(\ell_{ij}) + \sqrt{\mathcal{A}(\ell_{ij})^2 -1} \right) \,, \\
\mathcal{A}(\ell_{ij}) :&=  1 + 2 \, \frac{u_{h}^2}{u_{c}^2} \sinh\left( \frac{\ell_{ij}}{2 L^2 u_{h}}\sqrt{1 - \frac{u_{c}^2}{u_{h}^2}}  \right)^2,
\end{split}
\end{equation}
where $G$ is the Newton constant,  $u_c:=1/r_c$, and 
$u_{h}:=1/r_{h}$ is proportional  to the inverse temperature 
\begin{equation} \label{temp123}
 \beta =  2 \pi L^2 \, u_{h} \,.
\end{equation}
The length $\ell_{ij} := |\tilde{x}_{i}-\tilde{x}_{j}|/\sqrt{1 - u_{c}^2/u_{h}^2}$ corresponds to the length of the single interval $|x_{i}-x_{j}|$ in the dual field theory.   
 Eq.~\eqref{PTEE} reduces to the usual holographic entanglement entropy in \cite{Ryu:2006bv} as $u_{c} \rightarrow 0$. 

To compare this with the field theory result \eqref{dsasi}, let us consider a small deformation or cutoff ($u_{c} \ll u_{h}$)~\cite{Chen:2018eqk},
\begin{align} 
\begin{split} 
S^{H}(\ell_{ij}) 
&= \frac{L}{2G} \log \Biggl[\frac{2 u_{h}  \sinh \left(\frac{\ell_{\ij} }{2 L^2 \,u_{h} }\right)}{u_{c}}   \\
&\qquad\qquad\quad \times \left(1 - \frac{u_{c}^2}{4 L^2 u_{h}^3} \left(    \ell_{ij} \coth \left(\frac{\ell_{\ij} }{2 L^2 \, u_{h} }\right)  - \frac{ L^2 u_{h}}{\sinh \left(\frac{\ell_{\ij} }{2 L^2 \, u_{h} }\right)^2}      \right)   \right)      + \mathcal{O}\left(\frac{u_{c}^3}{u_{h}^3}\right)\Biggr]   \\
&\sim \frac{L}{2G} \log \left(\frac{2 u_{h}  \sinh \left(\frac{\ell_{\ij} }{2 L^2 \,u_{h} }\right)}{u_{c}}\right) -  \frac{u_{c}^2}{8 G L \, u_{h}^3} \left( \ell_{ij} \coth \left(\frac{\ell_{\ij} }{2 L^2 \, u_{h} }\right)  - \frac{L^2 u_{h}}{\sinh \left(\frac{\ell_{\ij} }{2 L^2 \, u_{h} }\right)^2}  \right)  \,.\label{}
\end{split}
\end{align}
By further considering the  ``high temperature'' limit ($ \beta = 2\pi L^2 \, u_{h} \ll \ell_{ij}$), that is to say, $u_{c} \ll u_{h} \ll \frac{\ell_{ij}}{2\pi L^2} $, the holographic entanglement entropy $S^{H}(\ell_{ij})$ in \eqref{PTEE} becomes 
\begin{align} 
\begin{split}
S^{H}_{\, \text{High T}}(\ell_{ij}) &\sim \frac{L}{2G} \log \left(\frac{2 u_{h}  \sinh \left(\frac{\ell_{\ij} }{2 L^2 \,u_{h} }\right)}{u_{c}}\right) -  \frac{\ell_{ij} \, u_{c}^2}{8 G L \, u_{h}^3}\coth \left(\frac{\ell_{\ij} }{2 L^2 \, u_{h} }\right)  \\
&= \frac{c}{3} \log \left(\frac{\beta  \sinh \left(\frac{\pi  \ell_{\ij} }{\beta }\right)}{\pi  \epsilon}\right) - \mu \, \frac{\pi ^4 c^2 \, \ell_{ij}}{9 \beta ^3}\coth \left(\frac{\pi  \ell_{\ij} }{\beta }\right)\,,\label{HighTcase2}
\end{split}
\end{align}
with~\cite{Chen:2018eqk, McGough:2016lol, Strominger:1997eq}
\begin{equation} \label{Relation}
\begin{split}
c = \frac{3L}{2G}\,, \quad \epsilon = L^2 u_{c} \,, \quad \mu = \frac{6 L^4}{\pi c}u_{c}^2   \,,
\end{split}
\end{equation}
where $c$ is the central charge and $\epsilon$ is the corresponding UV cutoff in the dual field theory.
Note that the second term of \eqref{HighTcase2} agrees with \eqref{dsasi}, i.e. the first order correction in $\mu$ to the holographic entanglement entropy {\it only at high temperature limit} ($ \beta \ll \ell_{ij}$) matches with the field theory result~\cite{Chen:2018eqk}. 

\subsection{Field theory: two intervals and phase transition}\label{field123}

In section \ref{sec32}, for two intervals, we find that there are {two phases of the entanglement entropy}: $s$-channel and $t$-channel. {In the 2d holographic CFT without the deformation, it has been shown that $\eta=1/2$ is the transition point between two channels~\cite{Hartman:2013mia} with the assumption that there are no other phases.}  We may ask what the effect of the small deformation on the phase transition is. Does it enhance the phase transition or not? To answer this question in the perturbed field theory we express the entanglement entropy of the deformed holographic CFT up to first order perturbation:
\begin{align} \label{FULL}
\begin{split}
S_{\,\text{s-ch}}(A) 
&= 
   \frac{c}{3} \log \left(\frac{\beta  \sinh \left(\frac{\pi  |x_{2}-x_{1}| }{\beta }\right)}{\pi  \epsilon}\right)
+ \frac{c}{3} \log \left(\frac{\beta  \sinh \left(\frac{\pi  |x_{4}-x_{3}| }{\beta }\right)}{\pi  \epsilon}\right)
+ \delta S_{\,\text{s-ch}}(A)    \,, \\
S_{\,\text{t-ch}}(A) 
&=  \quad x_2 \leftrightarrow  x_4  \quad \mathrm{and}\quad \text{s-ch}  \to \text{t-ch} \quad \mathrm{in} \quad
S_{\,\text{s-ch}}(A)   \,. 
\end{split}
\end{align}
Here, $S_{\,\text{s-ch}}(A)$ and $S_{\,\text{t-ch}}(A)$ are the entanglement entropy of the $s$-channel and $t$-channel up to first order perturbation, respectively, where $\delta S_{\,\text{s-ch}}(A)$ and $\delta S_{\,\text{t-ch}}(A)$ are the first order correction of the entanglement entropy given in \eqref{scds} and \eqref{tcds}, respectively.

{In the previous subsection we see that the field theory results match the holographic entanglement entropy in the high temperature limit: $\beta \ll \ell_{ij} = |x_{i}-x_{j}|$.} 
Thus, we expand \eqref{FULL} in terms of $\beta/ \ell_{ij} $
\begin{align} \label{stch11}
\begin{split}
S_{\,\text{s-ch}}(A)
& = \frac{c}{3} \left( \frac{\pi \left(|x_2-x_1|+|x_4-x_3|\right)}{\beta} - 2 \log\left(\frac{2\pi\epsilon}{\beta}\right) \right) - \mu \, \frac{\pi^4 c^2 \left(|x_2 - x_1| + |x_4 - x_3|\right)}{9 \beta^3} \,, \\
S_{\,\text{t-ch}}(A)
& = \quad x_2 \leftrightarrow  x_4  \quad  \mathrm{in} \quad
S_{\,\text{s-ch}}(A) \,,
\end{split}
\end{align}
where we used that $\sinh(\ell_{ij}/\beta) \sim \frac{e^{\ell_{ij}/\beta}}{2}$ and $\coth(\ell_{ij}/\beta) \sim 1$. Then we have 
\begin{align} \label{FIELDresult}
\begin{split}
S_{\,\text{s-ch}}(A) - S_{\,\text{t-ch}}(A)
& = -\frac{2 \, c \, \pi(x_3-x_2)}{3 \beta}\left( 1 - \mu \, \frac{c \, \pi^3 }{3 \beta^2}\right)   < 0 \,.
\end{split}
\end{align}
Because our field theory method is only reliable in the small $\mu/\beta^2$ regime, \eqref{FIELDresult} is always negative so there is no phase transition: the $s$-channel is always dominant in the high temperature. 

\subsection{Holography: two intervals (symmetric case)}

As we showed in the previous section, our field theory computation cannot capture any phase transition in its validity regime: the small deformation and high temperature regime. 
However, since the holographic entanglement entropy formula \eqref{PTEE} can be defined at any temperature (any $u_{h} > 0$ via \eqref{temp123}) and any cutoff $u_{c}<u_{h}$,\footnote{This inequality comes from the positive definite condition in  \eqref{PTEE}. } we can explore the transition in the whole region $0< u_{c} <u_{h}$. 
{Note that, from the field theory point of view, \eqref{PTEE} may not make sense for the entanglement entropy when $\ell_{ij}< u_c$ (See footnote 12). Therefore,  if the cutoff becomes bigger,  the {\it holographic} entanglement entropy may not be dual to the field theory entanglement entropy even though the {\it holographic} entanglement entropy is a well-defined object in gravitational language.
Because our purpose  in this section is to investigate the phase transition of the \textit{holographic} entanglement entropy (5.4)[56, 60] itself, our computation includes $\ell_{ij}< u_c$ case. However, at least we do not need to worry about this issue in the high temperature limit ($\ell_{ij} \gg u_{h}>u_{c}$).}

In the case of two intervals, we have two configurations of minimal surfaces as shown in Fig. \ref{hee}.
Then, the holographic entanglement entropy is chosen as the one having a smaller minimal surface:
\begin{equation} \label{kjhuy}
\begin{split}
S^{H} &=\text{min} \left\{ S_{\,\text{s-ch}}^{H} ,\,\,  S_{\,\text{t-ch}}^{H}  \right \}, 
\end{split}
\end{equation}
%
%
where 
\begin{align} \label{yui678}
\begin{split}
S_{\,\text{s-ch}}^{H}:=S^{H}(\ell_{12}) + S^{H}(\ell_{34}),& \qquad  S_{\,\text{t-ch}}^{H}:=S^{H}(\ell_{14}) + S^{H}(\ell_{23}) \,,
\end{split}
\end{align}
and here $S^{H}(\ell_{ij})$ is  the holographic entanglement entropy for the single interval $\ell_{ij} = |x_i-x_j|$. For convenience, we assume $x_1 < x_2 < x_3 < x_4$.

In summary, to find $S^H$  our task is i) to compute  $S_{\,\text{s-ch}}^{H}$ and $S_{\,\text{t-ch}}^{H}$ in \eqref{yui678} by using the single interval formula \eqref{PTEE}, ii) to compare them and pick up the one with a smaller value, which is \eqref{kjhuy}. 
In order to quantify the transition points we define
\begin{equation} \label{RatioEq}
\begin{split}
\mathcal{S}_{c} : = \frac{ S_{\,\text{s-ch}}^{H}  }  { S_{\,\text{t-ch}}^{H} } \,.
\end{split}
\end{equation}
Thus,  if $\mathcal{S}_{c} > 1$, $S^{H}  = S_{\,\text{t-ch}}^{H} $, while if $\mathcal{S}_{c} < 1$, $S^{H}  = S_{\,\text{s-ch}}^{H} $. The curves of $\mathcal{S}_{c} = 1$ are the transition points or the phase boundaries.  
Setting $\ell_{12}$ as a scaling parameter we deal with the following scaled parameters. 
\begin{align} \label{DEF}
\begin{split}
\bar{\ell}_{34}:= \frac{\ell_{34}}{\ell_{12}}, \quad  \bar{\ell}_{23}:= \frac{\ell_{23}}{\ell_{12}}, \quad \bar{L}:= \frac{L}{\ell_{12}}, \quad \bar{u}_{c}:= u_{c} \, \ell_{12}, \quad \bar{u}_{h}:= u_{h} \, \ell_{12}\,.
\end{split}
\end{align}
From here, we take $\bar{L} = 1$ without loss of generality.

\begin{figure}[]
 \centering
     \subfigure[$\bar{u}_{h}=2$. The red dot will be explained in Fig.~\ref{phastTa}.]
     {\includegraphics[width=7.2cm]{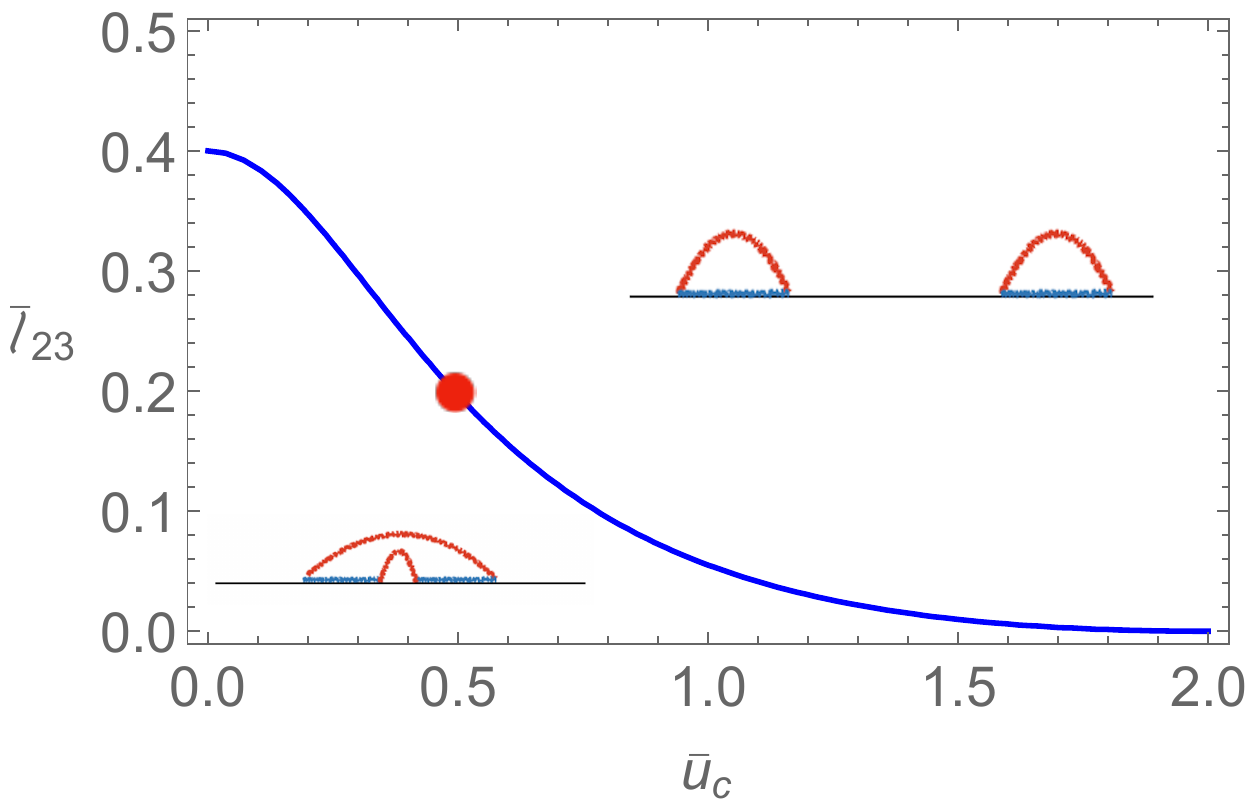} \label{LOWTEM}}
     \subfigure[$\bar{u}_{h}=2,\, 0.5,\, 0.1$]
     { \includegraphics[width=7.2cm]{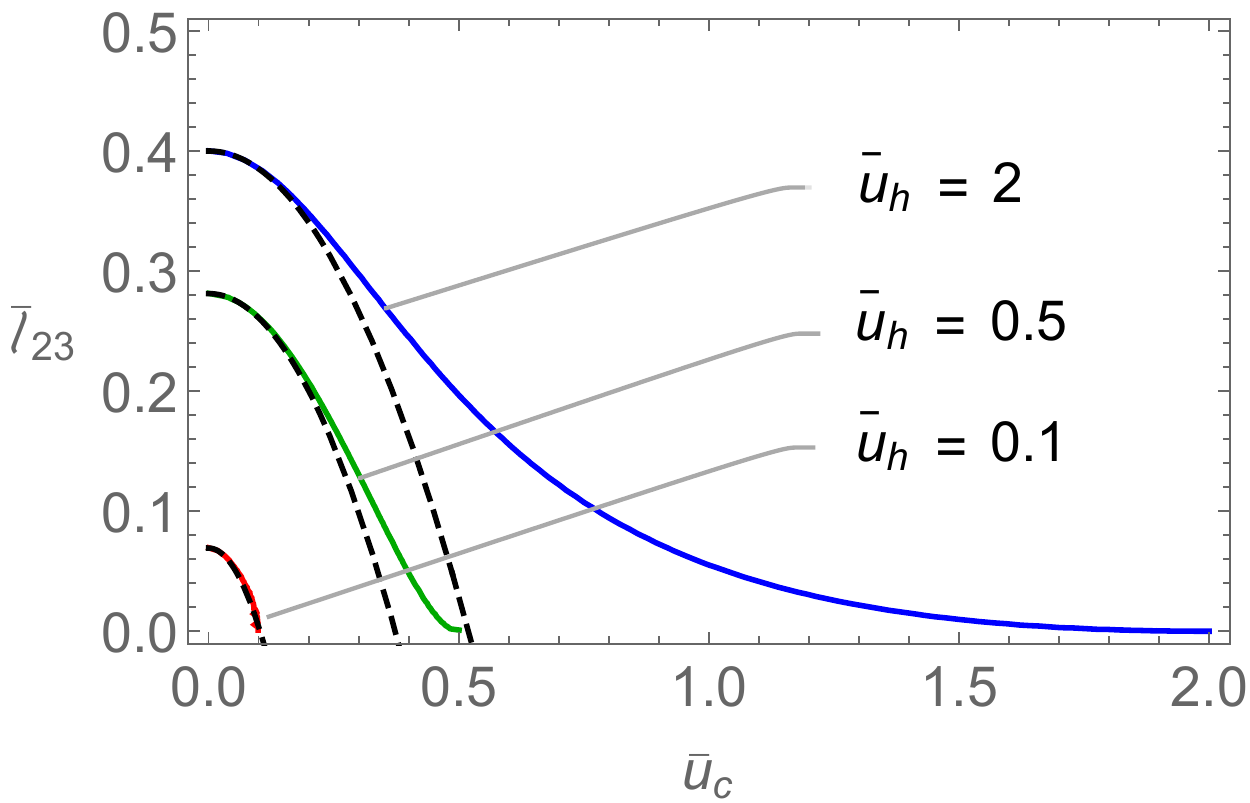} \label{AllTEM}}
          \caption{{Transition curves of symmetric case {($\bar{\ell}_{34}= \ell_{34}/\ell_{12} =1$), where $\bar{\ell}_{23} = \ell_{23}/\ell_{12}$, $\bar{u}_c = u_c \ell_{12}$ and  $\bar{u}_h = u_h \ell_{12}$.} All solid curves represent phase transition points satisfying $\mathcal{S}_{c} = 1$ (see \eqref{RatioEq}) with the various temperature $\bar{u}_h$, and the black dashed curves represent approximate formulas \eqref{Asymp}. The region above(below) the solid curves corresponds to the $s(t)$-channel.}} \label{TransitionPointFigure}
\end{figure}
Let us first consider a symmetric case ($\bar{\ell}_{34}= 1$).
Fig. \ref{TransitionPointFigure} shows the curves of $\mathcal{S}_{c} = 1$ (see \eqref{RatioEq}) in the plane of $\bar{\ell}_{23}$ and $\bar{u}_c$ at fixed temperature $\bar{u}_h$.  Fig. \ref{LOWTEM} is for $\bar{u}_h=2$ and  Fig. \ref{AllTEM} is for $\bar{u}_h=$2(blue), 0.5(green), 0.1(red). The region above(below) the solid curves correspond to the $s(t)$-channel.  The black dashed curves display the approximate results near $\bar{u}_{c} = 0$ in \eqref{Asymp}. 
From Fig. \ref{TransitionPointFigure} we find the followings. 

\begin{figure}[]
 \centering
     \subfigure[$\bar{\ell}_{23} > \bar{\ell}_{23}^{\,\,\,\, c}$]
     {\includegraphics[width=4.831cm]{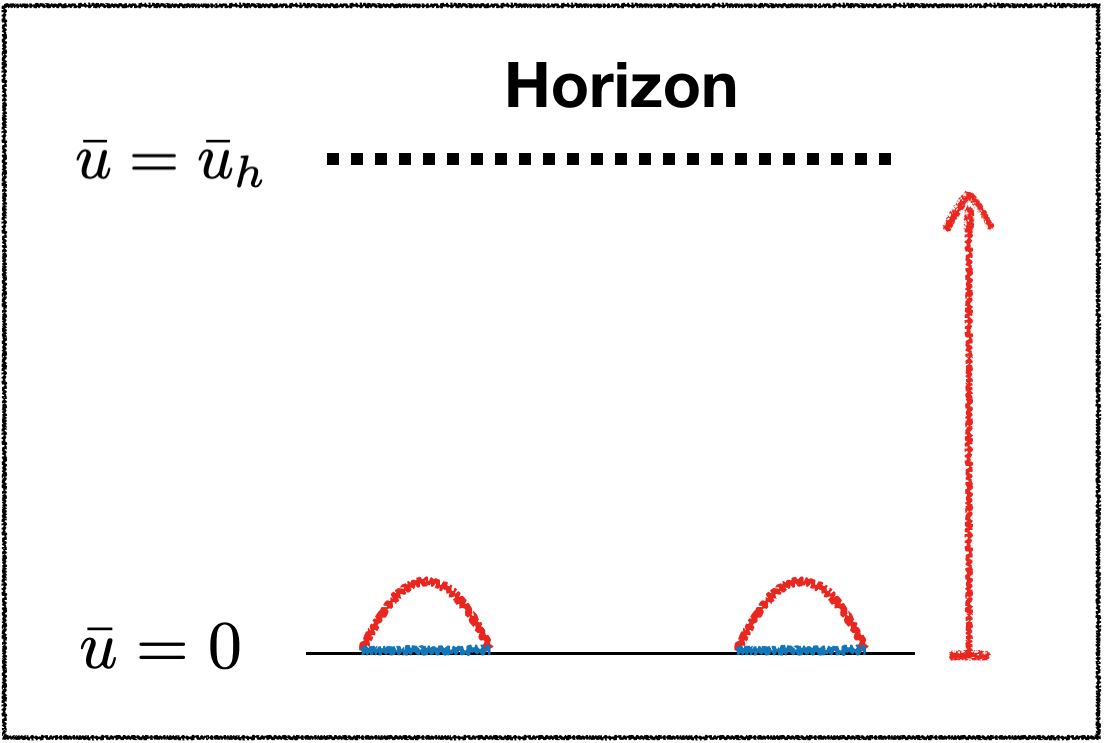} \label{TABLEFIG1}}
     \subfigure[$\bar{\ell}_{23} < \bar{\ell}_{23}^{\,\,\,\, c}$]
     {\includegraphics[width=4.831cm]{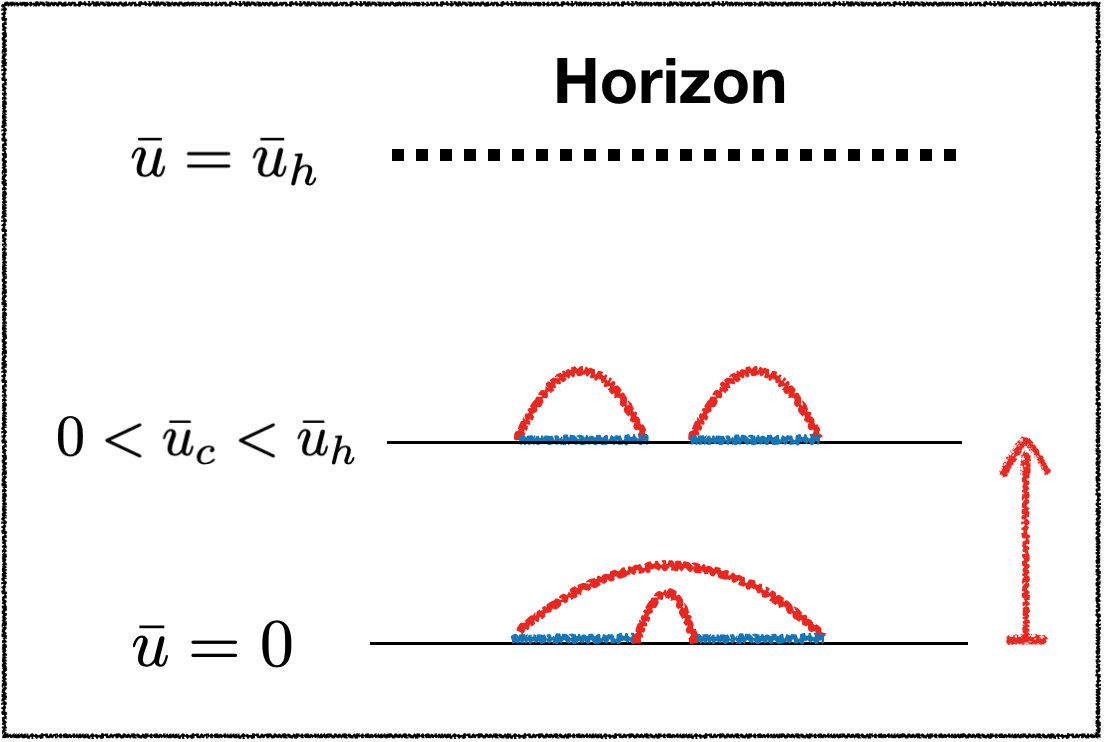} \label{TABLEFIG2}} 
     \subfigure[$\bar{\ell}_{23} \simeq 0$]
     {\includegraphics[width=4.831cm]{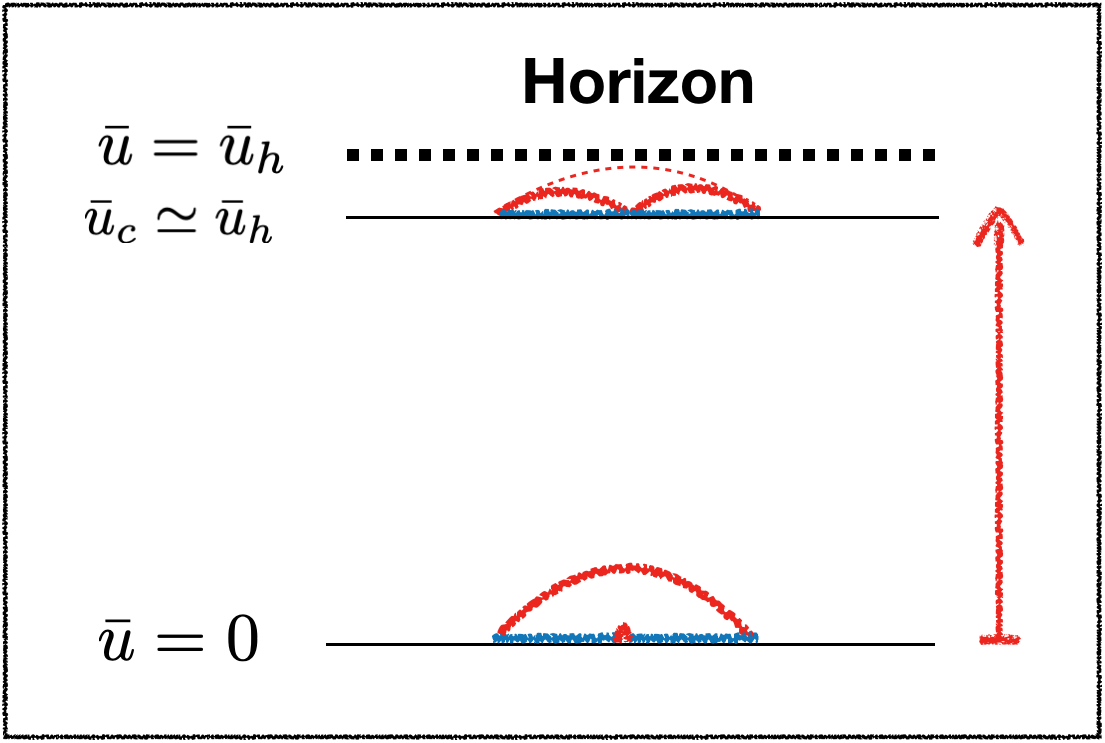} \label{TABLEFIG3}}
          \caption{{Cutoff ($\bar{u}_c$) dependence of the phase transition. 
(a) If the separation is big enough ($\bar{\ell}_{23} > \bar{\ell}_{23}^{\,\,\,\, c}$, see \eqref{Asymp}) there is no phase transition.  The $s$-channel is always favored. (b)(c) Otherwise, there is a phase transition from the $t$-channel to $s$-channel as the cutoff $\bar{u}_c$ increases.}} \label{TABLEFIG}
\end{figure}

\paragraph{Field theory results:} The perturbative field theory regime we studied in section \ref{field123} qualitatively corresponds to the left-upper corner of Fig. \ref{TransitionPointFigure}. Qualitatively speaking, `Left' corresponds to the small deformation and `up' corresponds to the high temperature limit\footnote{Rigorously speaking, the high temperature limit also includes $\bar{u}_h \ll 1$.}. The left-upper corner is always the $s$-channel, which is consistent with the field theory result. 
\paragraph{Separation dependence:} For fixed $\bar{u}_c$ and $\bar{u}_h$,  as the separation $\bar{\ell}_{23}$ increases, the $s$-channel is favored. Note that there is {\it always} a phase transition because $\bar{u}_c < \bar{u}_h$. 
\paragraph{Cutoff dependence:} Let us first define $\bar{\ell}_{23}^{(0)}$ by the maximum  $\bar{\ell}_{23}$ {for fixed  $\bar{u}_h$} allowing the phase transition. For example, $\bar{\ell}_{23}^{(0)} \approx 0.4$ in Fig. \ref{LOWTEM}. If $\bar{\ell}_{23} > \bar{\ell}_{23}^{(0)}$, there is no phase transition; always the $s$-channel is favored (Fig. \ref{TABLEFIG1}).  If $\bar{\ell}_{23} < \bar{\ell}_{23}^{(0)}$,  as the cutoff $\bar{u}_{c}$ increases, the $s$-channel is favored (Fig. \ref{TABLEFIG2}). Even for $\bar{\ell}_{23} \sim 0$ it undergoes a phase transition near $\bar{u}_c \sim \bar{u}_h$ (Fig. \ref{TABLEFIG3}).

\paragraph{Temperature dependence:}
Our result at low temperature is qualitatively consistent with the one at zero temperature~\cite{Ota:2019yfe}. In general, large separation and large cutoff favor the $s$-channel. There are some ranges  for the $t$-channel around the corner $\bar{u}_{c} \sim \bar{\ell}_{23} \sim 0$. However, as the temperature increases(blue to green to red in Fig. \ref{AllTEM}), this $t$-channel range shrinks towards $\bar{u}_{c} \sim \bar{\ell}_{23} \sim 0$, and finally vanishes for all cutoff  ($\bar{u}_{c}$) and the separation $\bar{\ell}_{23}$ at $\bar{u}_h = 0$.

\paragraph{Maximum separation for the phase transition:}

By expanding \eqref{RatioEq} in terms of $\bar{u}_{c} \ll 1$, we obtain an approximate formula $\bar{\ell}_{23}^c(\bar{u}_c, \bar{u}_h, \bar{\ell}_{34})$ for transition points $\mathcal{S}_{c} = 1$:
\begin{equation} \label{Asymp}
\begin{split}
\bar{\ell}_{23}^c(\bar{u}_c, \bar{u}_h,\bar{\ell}_{34}=1) \,=\, \bar{\ell}_{23}^{\,\,\,(0)}  +  \frac{  \bar{u}_{c}^2}{2 \, \bar{u}_{h}^2}\left(  \bar{\ell}_{23}^{\,\,\,(0)} + 1  + \frac{ 1 - 2\bar{u}_{h} - (1 + 2 \bar{u}_{h}) e^\frac{2}{ \bar{u}_{h}} }{\sqrt{\left(-1 + e^{\frac{1}{\bar{u}_{h}}}\right)^2 \left(1 + e^{\frac{2}{\bar{u}_{h}}}  \right)} }   \right) + \mathcal{O}\left(\bar{u}_{c}^4\right) \,,
\end{split} 
\end{equation}
where $\bar{\ell}_{23}^{\,\,\,(0)} = \bar{\ell}_{23}(0, \bar{u}_h, 1)$ is the value without a cutoff:
\begin{align} 
\bar{\ell}_{23}^{\,\,\,(0)} &= -2 +  \bar{u}_{h} \log \left[1 - e^{\frac{1}{\bar{u}_{h}}}  + e^{\frac{2}{\bar{u}_{h}}} + \sqrt{\left(   -1 + e^{\frac{1}{\bar{u}_{h}}}    \right)^2 \left( 1+e^{\frac{2}{\bar{u}_{h}}}    \right)} \right]   \,, \label{Asymp2} \\
&\sim\,
	\hspace*{-0cm}\begin{cases}
\sqrt{2}-1                     \,, \qquad\qquad       \left(\bar{u}_{h} \gg 1\right) \,,  \\
(\log 2) \, \bar{u}_{h}    \,,  \quad\qquad        \left(\bar{u}_{h} \ll 1\right) \,.
	\hspace*{-0cm}\end{cases}  \label{TempDep}
\end{align}
The asymptotic formula \eqref{Asymp} is shown as black dashed curves in Fig. \ref{TransitionPointFigure}.
At a given temperature,  $\bar{\ell}_{23}^{\,\,\,(0)}$  is the maximum separation which allows the $t$-channel. If two intervals are farther from each other than $\bar{\ell}_{23}^{\,\,\,(0)}$ only the $s$-channel is allowed. Note that $\bar{\ell}_{23}^{\,\,\,(0)} \sim 0.4$ at the zero temperature limit {($\bar{u}_{h} \to \infty$)}, which is already close to the case at $\bar{u}_h = 2$ in Fig. \ref{LOWTEM}. 

\subsection{Holography: two intervals (asymmetric case)}

\begin{figure}[]
\centering
\subfigure[$\bar{\ell}_{34} = 0.1$]
    { {\includegraphics[width=4.53cm]{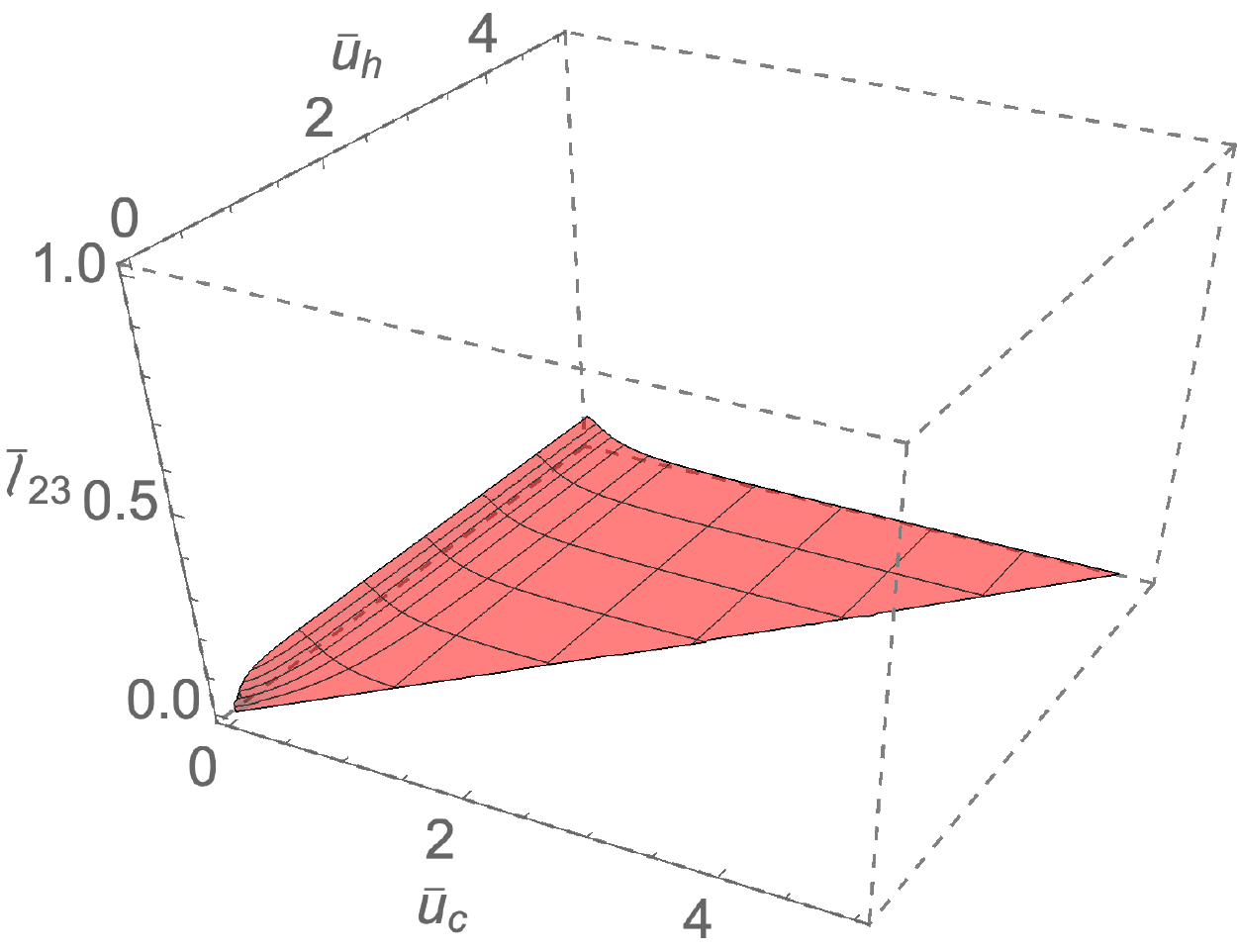}} \label{TransitionPointFigure2a}}
\subfigure[$\bar{\ell}_{34} = 1$]
    { {\includegraphics[width=4.53cm]{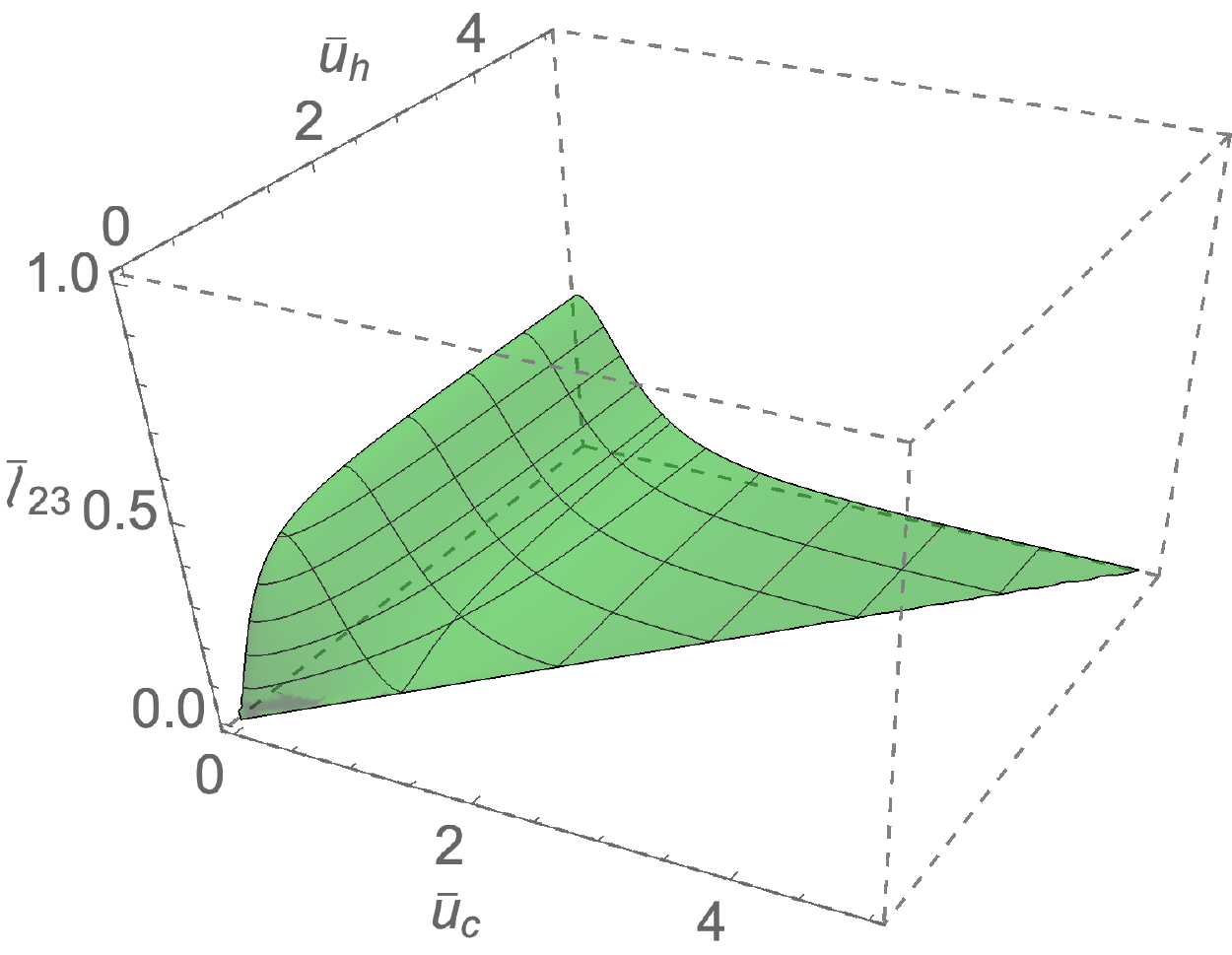}}\label{TransitionPointFigure2b}}
\subfigure[$\bar{\ell}_{34} = 10$]
   { {\includegraphics[width=4.53cm]{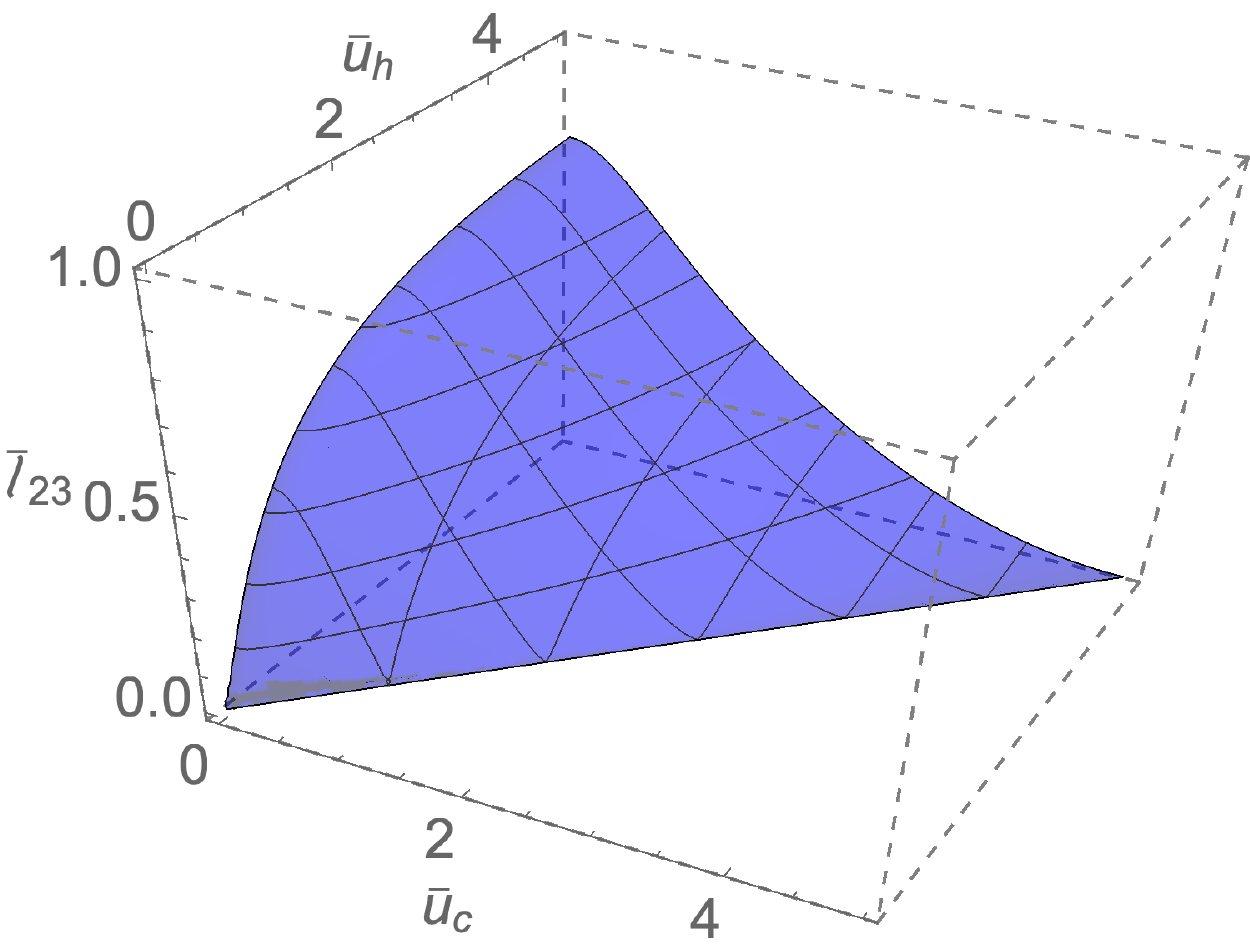}}\label{TransitionPointFigure2c}}
\caption{{The transition surface in ($\bar{u}_{c}, \bar{u}_{h}, \bar{\ell}_{23}$) space for $\bar{\ell}_{34} = 0.1,1,10$. The region above(below) the transition surface is for the $s$($t$)-channel. As $\bar{\ell}_{34}$ increases, the parameter region for the $t$-channel increases.}}\label{TransitionPointFigure2}
\end{figure}

Next, we consider the asymmetric case ($\bar{\ell}_{34} \neq 1$).
Fig. \ref{TransitionPointFigure2} shows the phase transition surface for $\bar{\ell}_{34} = 0.1, 1, 10$.  The region above the transition surface is for the $s$-channel.  Fig. \ref{TransitionPointFigure2b} in particular corresponds to the symmetric case ($\bar{\ell}_{34} \neq 1$) which is the three dimensional version of Fig. \ref{AllTEM}. As  $\bar{\ell}_{34}$ decreases, the $t$-channel is more suppressed. As  $\bar{\ell}_{34}$ increases, the parameter region for the $t$-channel increases and saturates  to the maximum region. 

\begin{figure}[]
\centering
     {\includegraphics[width=9cm]{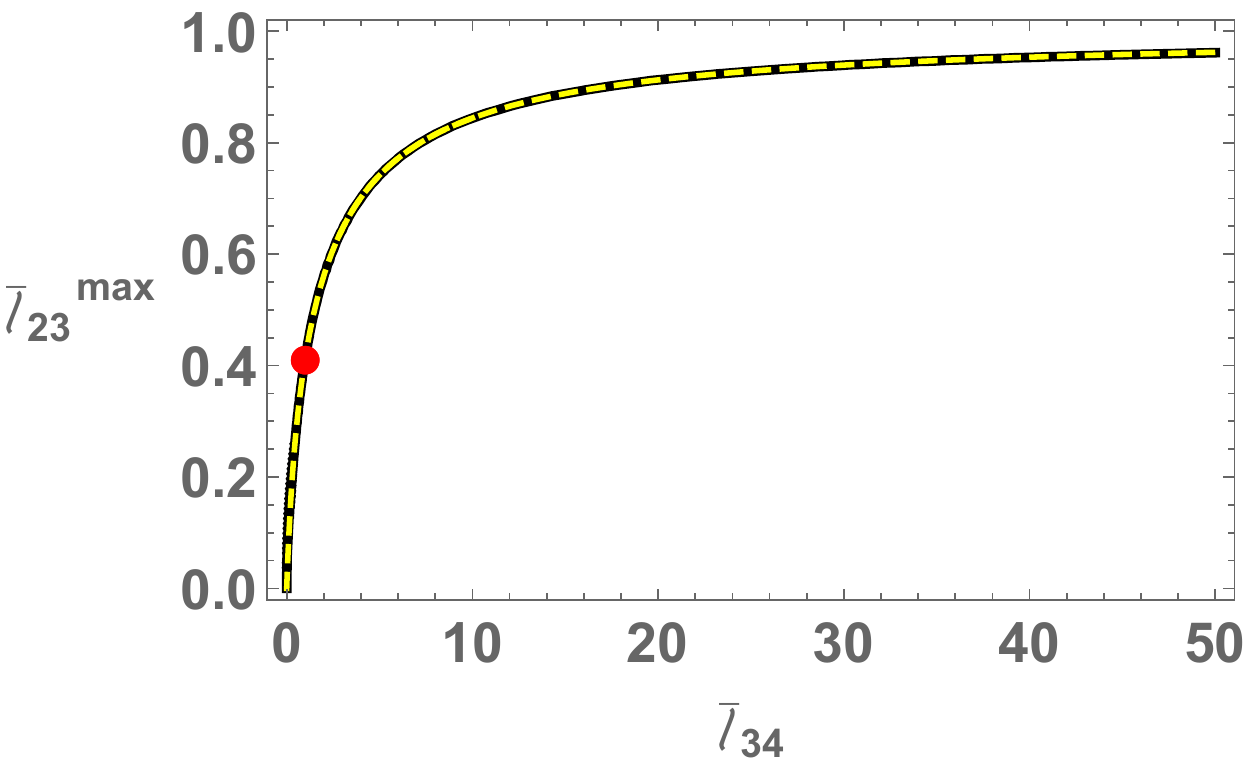}}
\caption{{$\bar{\ell}_{23}^{\,\,\,\, max}$ vs $\bar{\ell}_{34}$.   If the separation ($\bar{\ell}_{23}$) is bigger than $\bar{\ell}_{23}^{\,\,\,\, max}$, only the $s$-channel is available.
The black solid line is the numerical plot with ($\bar{u}_{c}=0$, $\bar{u}_{h}=100$). The yellow dashed line represents the analytic result in \eqref{MAXanal}. The red dot corresponds to the symmetric case ($\bar{\ell}_{34}=1$).} }\label{MAXVAL}
\end{figure}

\paragraph{Maximum separation for the phase transition:}
Similarly to the symmetric case, there is the maximum separation for the phase transition.  If the separation is bigger than this, only the $s$-channel is available.   From Fig. \ref{TransitionPointFigure2} we find that the transition point  at $\bar{u}_c=0$ at $\bar{u}_h \rightarrow \infty$ has the maximum separation.  By collecting these points for various $\bar{\ell}_{34}$ ($\bar{u}_{c}=0$, $\bar{u}_{h}=100$), we make a black curve in Fig. \ref{MAXVAL}. The maximum point increases as $\bar{\ell}_{34}$ increases, but it does not exceeds 1.
Indeed, this curve can be understood as follows.
Without a cutoff ($\bar{u}_c = 0$),  in the limit $\bar{u}_h \rightarrow \infty$, $\bar{\ell}_{23}$ satisfying $\mathcal{S}_{c} = 1$  saturates to 
\begin{align} \label{MAXanal}
\begin{split}
\bar{\ell}_{23}^{c}(0,\infty,\bar{\ell}_{34})   \,=\,     \frac{-1 - \bar{\ell}_{34} + \sqrt{(1+\bar{\ell}_{34})^2 + 4 \, \bar{\ell}_{34} }}{2}  \,,
\end{split}
\end{align}
which is plotted as the yellow dashed curve in Fig. \ref{MAXVAL}.  Note that 
\begin{equation}
\bar{\ell}_{23}^{c}(0,\infty,\bar{\ell}_{34})   \,\rightarrow 1  \,,\label{lmax1}
\end{equation}
as $\bar{\ell}_{34} \rightarrow \infty $. 
Alternatively, we can also understand \eqref{MAXanal} by the cross ratio $\eta$. This $\eta$ is defined in the 2d CFT, and in this case the phase transition occurs at $\eta = 1/2$~\cite{Hartman:2013mia, Headrick:2010zt}\footnote{In higher dimensional cases, the phase transition of the holographic entanglement entropy of two strips at finite temperature was also studied \cite{Fischler:2012uv, BabaeiVelni:2019pkw}.}.   Since we are considering the case $\bar{u}_c = 0$ and $\bar{u}_h \rightarrow \infty$, we may use that criteria i.e.
\begin{align} \label{ETARELATION}
\begin{split}
\eta = \frac{(z_1-z_2)(z_3-z_4)}{(z_1-z_3)(z_2-z_4)} 
        = \frac{(x_1-x_2)(x_3-x_4)}{(x_1-x_3)(x_2-x_4)}                                   
       = \frac{\bar{\ell}_{34}}{(1+\bar{\ell}_{23})(\bar{\ell}_{23}+\bar{\ell}_{34})} = \frac{1}{2} \,,
\end{split}
\end{align}
which implies \eqref{MAXanal}.

\begin{figure}[]
 \centering
     \subfigure[First order phase transition]
     {\includegraphics[width=7.2cm]{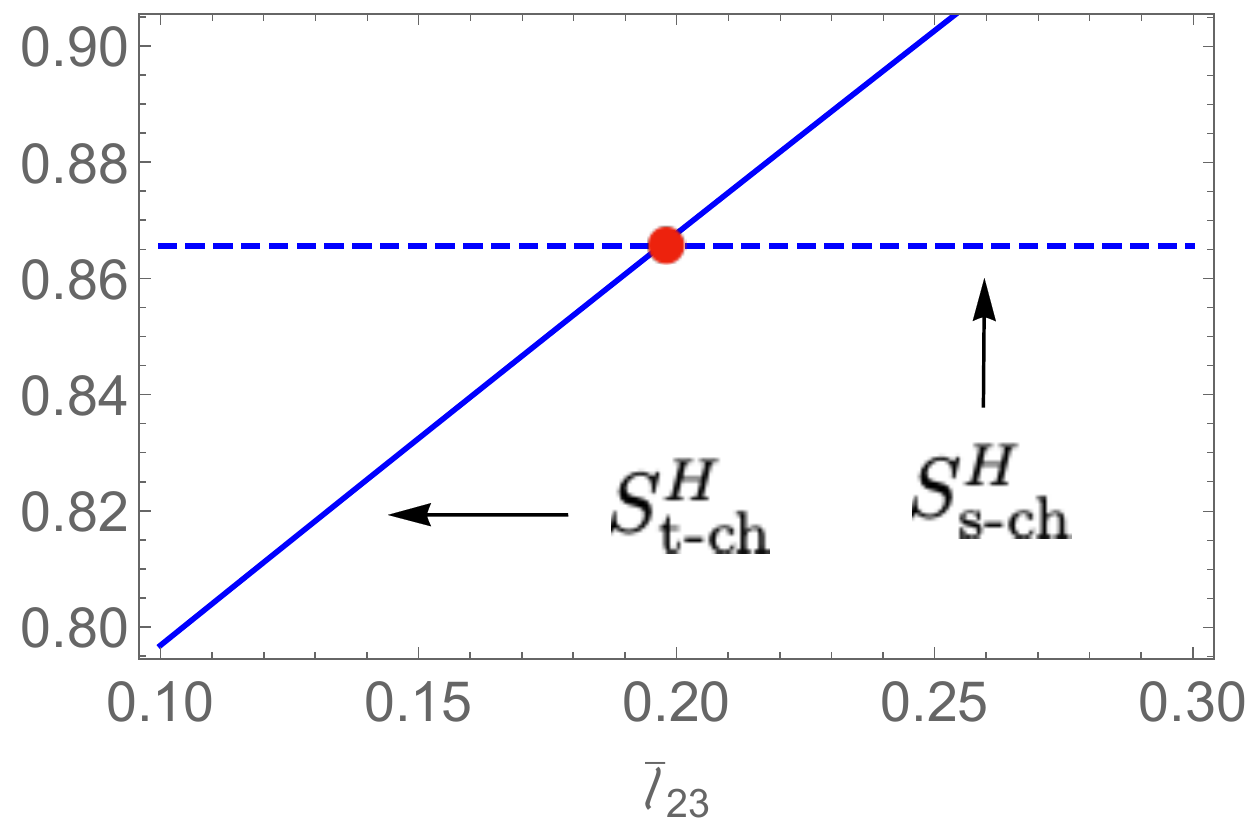} \label{phastTa}}
     \subfigure[(Schematic) second order phase transition]
     { \includegraphics[width=7.2cm]{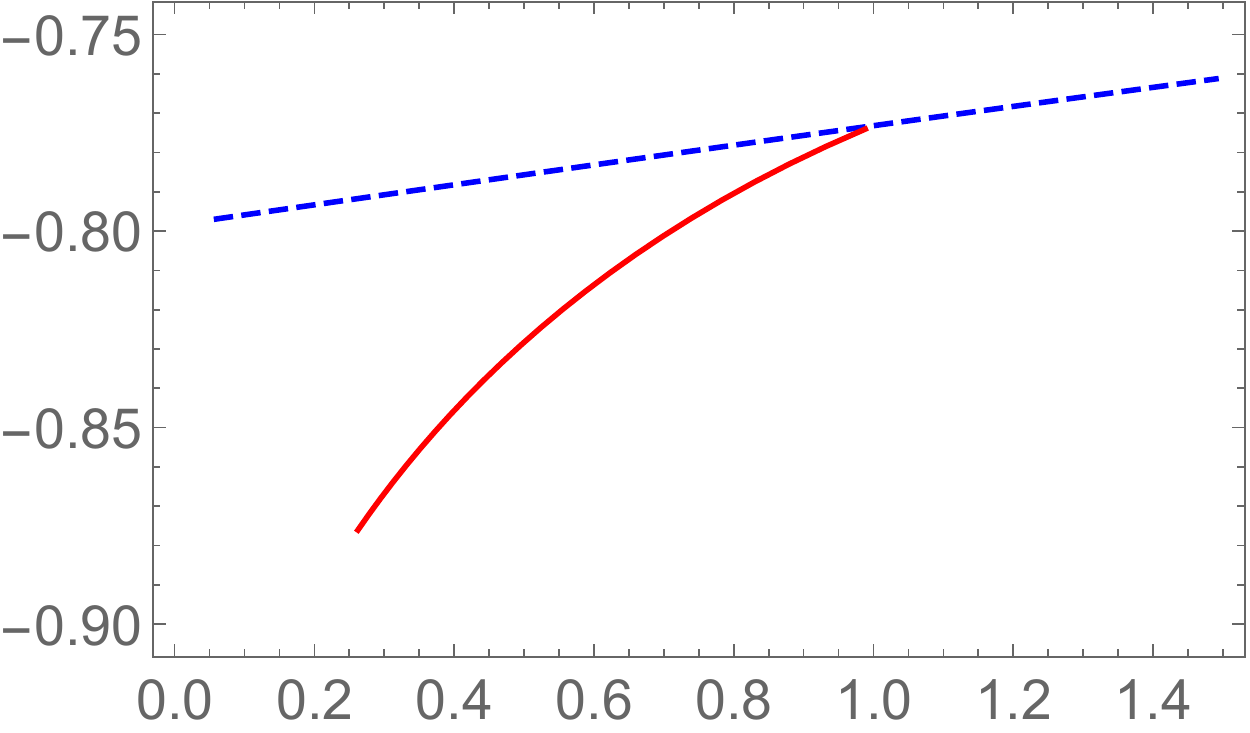} \label{phastTb}}
          \caption{Intuitive understanding of phase transitions} \label{phastT}
\end{figure}

{
All phase transitions in Fig. \ref{TransitionPointFigure2} is the first order phase transition. For example, let us consider the phase transition at $\bar{u}_h=2$ and $\bar{u}_c=0.5$. See the red dot in Fig.~\ref{LOWTEM}. As we increase $\bar{\ell}_{23}$ the configuration of smaller area changes from the $t$-channel to $s$-channel. 
It can be seen concretely in Fig.~\ref{phastTa}. As $\bar{\ell}_{23}$ increases, the entanglement entropy of the $s$-channel ($S^H_\text{s-ch}$) is constant (dotted line) because the entanglement surface does not change, while the entanglement entropy of the $t$-channel ($S^H_\text{t-ch}$) monotonically increases (solid line) because the entanglement surface becomes bigger. The same argument applies to all transition points in Fig.~\ref{TransitionPointFigure2} so it is natural to have the first order phase transition.

Intuitively, the first order phase transition is natural if two configurations are available for all parameter range as in Fig.~\ref{phastTa}.  In this case, there will be a cross point of two curves at the phase transition, so the first derivative at that point will be different (`first' order transition) in general. To have a second order (or continuous) phase transition, usually we have one configurations before the phase transition occurs and two configurations are available after the phase transition. See the schematic picture in Fig.~\ref{phastTb}, where the blue dotted curve goes to the red curve after the phase transition. A good example of this type of phase transition is a holographic superconductor~\cite{Gubser:2008px, Hartnoll:2008vx}.
}

\section{Conclusions} \label{summary}

In this work we have studied the entanglement entropy and the R\'{e}nyi entropy of multiple intervals in 2d CFT at finite temperature with the first order perturbation by the $T\overline{T}$ deformation. 

To compute the R\'{e}nyi entropy, computations of the correlation functions between the twist operators and $T\overline{T}$ are crucial. 
We have derived the general formula to compute the R\'{e}nyi entropy (also entanglement entropy as a special case of the R\'{e}nyi entropy) in general CFT up to the first {order} deformation. 

By using this formula we have found that the entanglement entropy of multiple intervals in the deformed holographic CFT is the sum of the one of a single interval. This is a non-trivial result from the field theory side while it looks straightforward from a holographic viewpoint via the Ryu-Takayanagi formula. In other words, it provides a non-trivial consistency check of holography with the $T\overline{T}$ deformation. 

On the contrary, the R\'{e}nyi  entropy of two intervals is the sum of the one of a single interval only if the distance between two intervals is large enough. It can be intuitively understood by the fact that the holographic  R\'{e}nyi  entropy are related with the cosmic brane which has a tension, contrary to the Ryu-Takayanagi surface. Thus, in general there will be a back-reaction, which will disappear only if two intervals are far away.

{Moreover, we also have found an interesting observation when the cross ratio goes to unity. We show that in this case there is a finite mixing effect between the holomorphic and anti-holomorphic parts to the R\'{e}nyi entropy which vanishes for the entanglement entropy but not for the R\'{e}nyi entropy. Because of this mixing, $\delta S_n(A)$ of the two intervals does not become the sum of the single interval's. We provide arguments to understand this non-vanishing mixing effect from the perspective of field theory and holography.

In general, if $\eta \rightarrow 0$, there is no mixing term and the cluster decomposition (such as the factorization of the CFT correlation function) is allowed, while if $\eta \rightarrow 1$ there should be a mixing term and the cluster decomposition is not allowed. In field theories with the conformal symmetry, the cluster decomposition is valid also for $\eta\to1$ by the conformal map so  there is no mixing term. However if the $T\overline{T}$ deformation is considered, conformal symmetry is broken (i.e., we cannot use a conformal map) so that the cluster decomposition property may not work in $\eta\to1$. Thus the mixing term may be non-zero. Our results can be considered as an example of this case: considering the finite $\mu$, mixing term in $\eta\to1$ is zero when $n=1$(the entanglement entropy) but finite when $n\neq1$(the R\'{e}nyi entropy). 

In holography, we can use a holographic prescription of the R\'{e}nyi entropy to understand mixing effect in $\eta\to1$ case; area of cosmic branes. 
If we interpret that the mixing term is from the interaction between two cosmic branes through the back-reaction ($n\neq1$), we can argue that the mixing effect may remain with $\mu\neq0$ (finite cutoff) as follows. 
i) If the distance between two cosmic branes are very far from each other such as the case with $\eta \rightarrow 0$, the interaction between them would be negligible (vanishing mixing term).
ii) However, in the $\eta\to1$ case, the distance between the two branes are always finite because one will shrink to the point on the finite radius cutoff (which is the definition of $\eta\to1$) and the other will be bounded by the horizon. Thus, we may expect that there will be a remaining interaction (non-vanishing mixing term). {This mixing effect will vanish if the cut-off or temperature goes to zero.}}

For two intervals, there are two configurations for the entanglement entropy, the so called $s$-channel and $t$-channel. They correspond to the disconnected Ryu-Takayangi surface and the connected surface respectively in holography.  Mathematically, both are available, but the entanglement entropy corresponds to the one with the smaller value. 
From our field theory computation, we have shown that the $s$-channel is always favored in the {\it small} deformation if the lengths of the intervals ($\ell_{12}$ and $\ell_{34}$) and the separation between them (${\ell}_{23}$) are much bigger than the temperature (${u}_h$), i.e. ${\ell}_{ij} \gg {u}_h \sim \beta$.  From our holography computation, we have confirmed it and, furthermore, shown that it is true also in the {\it large} (arbitrary) deformation.

Holographic framework can deal with an arbitrary deformation and temperature, contrary to the field theory method. By taking this advantage, we explored the whole parameter space of the deformation and temperature to identify the parameter range for $the t$-channel. 
We find that at a given deformation and temperature and $\ell_{34}$,  
if $\bar{\ell}_{23} = \ell_{23}/\ell_{12}$ becomes smaller there will be a phase transition from the $s$-channel to $t$-channel at some critical length, say $\bar{\ell}_{23}^c$. The critical length  $\bar{\ell}_{23}^c$ increases as the temperature or the deformation parameter decreases or $\ell_{34}$ increases. Thus, we find that the maximum value of $\bar{\ell}_{23}^c$ is determined by CFT(zero deformation) at zero temperature and ${\ell}_{34} \rightarrow \infty$, which is $\bar{\ell}_{23}^c \rightarrow 1$.

For the R\'{e}nyi  entropy of two intervals in this paper, we focused on the case $\eta \to 0$ or $\eta \to 1$ to use the factorization property of the four point function. However, in principle, it is possible to consider an arbitrary $\eta$ in the holographic CFT. 
Comparing it with the holographic mutual R\'{e}nyi  information {at least to first order in $n-1$ \cite{Dong:2016fnf}} may serve as another important consistency check of holography.
It will be also interesting to generalize the formalism for the R\'{e}nyi  entropy in this paper for other entanglement measures (for example, \cite{Calabrese:2012ew, Tamaoka:2018ned, Caputa:2018xuf, Dutta:2019gen}) by considering the suitable twist operators.
We leave these as future work.

\acknowledgments

We would like to thank Toshihiro Ota for fruitful discussions.
This work was supported by Basic Science Research Program through the National Research Foundation of Korea(NRF) funded by the Ministry of Science, ICT $\&$ Future Planning(NRF- 2017R1A2B4004810) and GIST Research Institute(GRI) grant funded by the GIST in 2019.

\appendix
\section{The integrals}  \label{appendixA}
In this Appendix, we perform the integrations \eqref{dsasi} and \eqref{mix987}.  Without loss of generality, we take $z_k$ are real and satisfy $z_1<z_2<z_3<z_4$.

\subsection{Correction to the entanglement entropy: $\delta S(A)$}   \label{appendixAa}
Let us first consider \eqref{dsasi}.
We will follow the same procedure as in~\cite{Chen:2018eqk}\footnote{This analysis can be taken as a generalization of~\cite{Chen:2018eqk} in the sense that we are considering an arbitrary $z_{i}$. If we choose $z_{1}=1$ and $z_{2} =e^{\frac{2\pi l }{\beta}}$, it reduces to~\cite{Chen:2018eqk}.}.
In the Euclidean world sheet ($x, \tau$), the integration reads
\begin{equation} \label{SingleInt}
\begin{split}
    \int_{\mathcal{M}} \dd^{2} w  & \,  \frac{z^2 (z_{1} - z_{2})^{2}}{(z - z_{1})^2(z-z_{2})^2}    \\
& = \int_{-\infty}^{\infty} \dd x \int_{0}^{\beta} \dd \tau \, \frac{e^{\frac{4 \pi (x+i \tau)}{\beta}} \left(z_{1} - z_{2} \right)^{2}}{ \left(\expo - z_{1}\right)^2 \left(\expo -z_{2} \right)^2} \,. \\
\end{split}
\end{equation}
We first integrate with respect to $\tau$.  The indefinite integral over $\tau$ is
\begin{equation} \label{taufirst}
\begin{split}
\frac{   i \beta \left[ \frac{z_{1}(z_{1}-z_{2})}{\expo - z_{1}}   +  \frac{z_{2}(z_{1}-z_{2})}{\expo - z_{2}}   + (z_{1}+z_{2})\left(\ln(\expo - z_{1}) - \ln(\expo-z_{2})\right)      \right]          }{ 2\pi (z_{1} - z_{2}) } \,.
\end{split}
\end{equation}
Because the first two terms in the numerator of \eqref{taufirst}
\begin{equation} \label{}
\begin{split}
 \frac{z_{1}(z_{1}-z_{2})}{\expo - z_{1}}   +  \frac{z_{2}(z_{1}-z_{2})}{\expo - z_{2}} \,,
\end{split}
\end{equation}
produce the same value as we put $\tau=0$ and $\tau=\beta$, these terms do not contribute to the result. 
Therefore, we only focus on the  logarithmic functions in \eqref{taufirst} which needs a careful analysis due to the branch cut.
We have the following identity useful for the logarithmic functions with the complex variables:
\begin{equation} \label{lnResult}
\begin{split}
\ln(\expo - z_{k}) \, \Big|_{\tau=0}^{\tau=\beta} \,=\, 2 \pi i,  \qquad  \left(e^{\frac{2 \pi x}{\beta}} > z_{k} \, \leftrightarrow \,  x > \frac{\beta}{2\pi} \ln z_{k} \right) \,.
\end{split}
\end{equation}
%
By \eqref{lnResult}, the logarithmic function in \eqref{taufirst} becomes
\begin{equation} \label{LogResult}
\begin{split}
\ln(\expo - z_{1}) - \ln(\expo-z_{2}) \, \Big|_{\tau=0}^{\tau=\beta} \,=\, 2 \pi i,  \quad  \left(\frac{\beta}{2\pi} \ln z_{1} < x < \frac{\beta}{2\pi} \ln z_{2} \right) \,.
\end{split}
\end{equation}
Therefore, with \eqref{LogResult},  the indefinite integral \eqref{taufirst} for $\tau=0$ and $\tau=\beta$ gives a $x$-independent result in the restricted range of $x$ :
\begin{equation} \label{}
\begin{split}
-\beta \, \frac{z_{1} + z_{2}}{z_{1} - z_{2}}, \qquad  \left(\frac{\beta}{2\pi} \ln z_{1} < x < \frac{\beta}{2\pi} \ln z_{2} \right) \,.
\end{split}
\end{equation}
In conclusion, the integral in \eqref{SingleInt} yields
%
\begin{equation} \label{SingleFinite}
\begin{split}
\int_{\mathcal{M}} \dd^{2} w  \,  \frac{z^2 (z_{1} - z_{2})^{2}}{(z - z_{1})^2(z-z_{2})^2} 
& = \int_{\frac{\beta}{2\pi} \ln z_{1}}^{\frac{\beta}{2\pi} \ln z_{2}} \dd x  \, \left(-\beta \, \frac{z_{1} + z_{2}}{z_{1} - z_{2}}\right) \\
& =  \frac{\beta^2}{2\pi} \, \frac{z_{1} + z_{2}}{z_{1} - z_{2}}  \ln\left( \frac{z_{1}}{z_{2}} \right) \,.
\end{split}
\end{equation}

\subsection{Correction to the R\'{e}nyi entropy: $\delta S_n(A)$}
Next let us consider \eqref{mix987}.
The integration is now written in the Euclidean world sheet ($x, \tau$) as
\begin{equation} \label{RenyiInt}
\begin{split}
&\int_{\mathcal{M}} \dd^{2} w   \frac{z^2 (z_{1} - z_{2})^{2}}{(z - z_{1})^2(z-z_{2})^2} \frac{\bar{z}^2 (\bar{z}_{3} - \bar{z}_{4})^{2}}{(\bar{z} - \bar{z}_{3})^2(\bar{z}-\bar{z}_{4})^2}    \\
& = \int_{-\infty}^{\infty} \dd x \int_{0}^{\beta} \dd \tau \, \frac{e^{\frac{8 \pi x}{\beta} } \left(z_{1} - z_{2} \right)^{2}\left(\bar{z}_{3} - \bar{z}_{4} \right)^{2}}{ \left(\expo - z_{1}\right)^2 \left(\expo -z_{2} \right)^2  \left(\expoo - \bar{z}_{3}\right)^2 \left(\expoo - \bar{z}_{4} \right)^2} \,. \\
\end{split}
\end{equation}
We follow the same procedure used in \eqref{appendixAa}. 
First, we perform the indefinite integral with respect to $\tau$:
\begin{align} \label{}
\begin{split}
&\int \dd \tau \, \frac{e^{\frac{8 \pi x}{\beta} } \left(z_{1} - z_{2} \right)^{2}\left(\bar{z}_{3} - \bar{z}_{4} \right)^{2}}{ \left(\expo - z_{1}\right)^2 \left(\expo -z_{2} \right)^2  \left(\expoo - \bar{z}_{3}\right)^2 \left(\expoo - \bar{z}_{4} \right)^2}\notag\\
=&\frac{-i \, \beta \, e^{\frac{8 \pi x}{\beta}}}{2\pi}  (z_{1}-z_{2})^2 (\bar{z}_{3}-\bar{z}_{4})^2  \left( \mathcal{A} + \mathcal{B} \right),
\end{split}
\end{align}
where $\mathcal{A}$ and $\mathcal{B}$ are given as follows,
\small
\begin{equation} \label{}
\begin{split}
\mathcal{A} =  &- \frac{
\frac{ z_{1}^3   }{\left( e^{\frac{2\pi(x+i\tau)}{\beta}} - z_{1} \right)   \left( e^{\frac{4 \pi x}{\beta}} - z_{1}\bar{z}_{3} \right)^2  \left( e^{\frac{4 \pi x}{\beta}} - z_{1}\bar{z}_{4} \right)^2     }    +  
\frac{ z_{2}^3   }{ \left( e^{\frac{2\pi(x+i\tau)}{\beta}} - z_{2} \right)   \left( e^{\frac{4 \pi x}{\beta}} - z_{2}\bar{z}_{3} \right)^2  \left( e^{\frac{4 \pi x}{\beta}} - z_{2}\bar{z}_{4} \right)^2     } 
}{(z_{1}-z_{2})^2}    \\
                        &+ \frac{
\frac{ \bar{z}_{3}^2   }{\left( e^{\frac{2\pi x}{\beta}} - e^{\frac{2\pi i \tau}{\beta}} \bar{z}_{3} \right)   \left( e^{\frac{4 \pi x}{\beta}} - z_{1}\bar{z}_{3} \right)^2  \left( e^{\frac{4 \pi x}{\beta}} - z_{2}\bar{z}_{3} \right)^2     }    +  
\frac{ \bar{z}_{4}^2   }{ \left(  e^{\frac{2\pi x}{\beta}} - e^{\frac{2\pi i \tau}{\beta}} \bar{z}_{4} \right)   \left( e^{\frac{4 \pi x}{\beta}} - z_{1}\bar{z}_{4} \right)^2  \left( e^{\frac{4 \pi x}{\beta}} - z_{2}\bar{z}_{4} \right)^2     } 
}{(\bar{z}_{3}-\bar{z}_{4})^2} e^{\frac{2 \pi x}{\beta}},  \\
\\
\mathcal{B} =  
& -\frac{1}{(z_{1}-z_{2})^3} \left(  \mathcal{B}_{1} + \mathcal{B}_{2} \right)  +  \frac{1}{(\bar{z}_{3}-\bar{z}_{4})^3} \left(  \mathcal{B}_{3} + \mathcal{B}_{4} \right) \,, \\
&\mathcal{B}_{1} = 
\frac{  z_{1}^2 \left(  z_{1}^2 (3 z_{1} - z_{2})\bar{z}_{3} \bar{z}_{4} - e^{\frac{4\pi x}{\beta}} z_{1} (z_{1}+z_{2})(\bar{z}_{3}+\bar{z}_{4})  - e^{\frac{8\pi x}{\beta}} (z_{1} - 3 z_{2}) \right)     }{ \left( e^{\frac{4 \pi x}{\beta}} - z_{1} \bar{z}_{3}  \right)^3 \left( e^{\frac{4 \pi x}{\beta}} - z_{1} \bar{z}_{4}  \right)^3 } \ln\left(e^{\frac{2\pi(x+i \tau)}{\beta}}  - z_{1} \right), \\
&\mathcal{B}_{2} = 
\frac{  z_{2}^2 \left(  z_{2}^2 (z_{1} - 3z_{2})\bar{z}_{3} \bar{z}_{4} + e^{\frac{4\pi x}{\beta}} z_{2} (z_{1}+z_{2})(\bar{z}_{3}+\bar{z}_{4})  - e^{\frac{8\pi x}{\beta}} (3z_{1} - z_{2}) \right)   }{ \left( e^{\frac{4 \pi x}{\beta}} - z_{2} \bar{z}_{3}  \right)^3 \left( e^{\frac{4 \pi x}{\beta}} - z_{2} \bar{z}_{4}  \right)^3 }   \ln\left(e^{\frac{2\pi(x+i \tau)}{\beta}}  - z_{2} \right),      \\
&\mathcal{B}_{3} = 
\frac{  \bar{z}_{3}^2 \left(  \bar{z}_{3}^2 (3 \bar{z}_{3} - \bar{z}_{4})z_{1} z_{2} - e^{\frac{4\pi x}{\beta}} \bar{z}_{3} (z_{1}+z_{2})(\bar{z}_{3}+\bar{z}_{4})  - e^{\frac{8\pi x}{\beta}} (\bar{z}_{3} - 3 \bar{z}_{4}) \right)     }{ \left( e^{\frac{4 \pi x}{\beta}} - z_{1} \bar{z}_{3}  \right)^3 \left( e^{\frac{4 \pi x}{\beta}} - z_{2} \bar{z}_{3}  \right)^3 } \ln\left(e^{\frac{2\pi x}{\beta}}  - e^{\frac{2\pi i \tau}{\beta}} \bar{z}_{3} \right),      \\
&\mathcal{B}_{4} = 
\frac{  \bar{z}_{4}^2 \left(  \bar{z}_{4}^2 (\bar{z}_{3} - 3 \bar{z}_{4})z_{1} z_{2} + e^{\frac{4\pi x}{\beta}} \bar{z}_{4} (z_{1}+z_{2})(\bar{z}_{3}+\bar{z}_{4})  - e^{\frac{8\pi x}{\beta}} (3 \bar{z}_{3} -  \bar{z}_{4}) \right)     }{ \left( e^{\frac{4 \pi x}{\beta}} - z_{1} \bar{z}_{4}  \right)^3 \left( e^{\frac{4 \pi x}{\beta}} - z_{2} \bar{z}_{4}  \right)^3 } \ln\left(e^{\frac{2\pi x}{\beta}}  - e^{\frac{2\pi i \tau}{\beta}} \bar{z}_{4} \right).       \\
\end{split}
\end{equation}
\normalsize
When we plug $\tau=0$ and $\tau=\beta$ into $\mathcal{A}$, they gives us the same value.
So, $\mathcal{A}$ does not contribute to the result.
In other words,
\begin{equation} \label{}
\begin{split}
\mathcal{A} \, \Big|_{\tau=0}^{\tau=\beta} \,=\, 0 \,.
\end{split}
\end{equation}
On the other hand, $\mathcal{B}$ has a logarithmic function which should be calculated carefully.
Because $\mathcal{B}$ has four different logarithmic functions in ($\mathcal{B}_{1}$, $\mathcal{B}_{2}$, $\mathcal{B}_{3}$, $\mathcal{B}_{4}$), we will have four modified integration ranges of $x$ after doing integration over $\tau$.
By using \eqref{lnResult}, the logarithmic functions in ($\mathcal{B}_{1}$, $\mathcal{B}_{2}$, $\mathcal{B}_{3}$, $\mathcal{B}_{4}$) will be substituted with $2 \pi i$ together with the following change of integration range of $x$:
\begin{equation} \label{xRange}
\begin{split}
   &  \mathcal{B}_{1}:  \int_{-\infty}^{\infty} \dd x    \,\,\rightarrow\,\,    \int_{\frac{\beta}{2\pi}\ln z_{1}}^{\infty} \dd x   \,, \qquad 
\,\,\,  \mathcal{B}_{2}:  \int_{-\infty}^{\infty} \dd x    \,\,\rightarrow\,\,    \int_{\frac{\beta}{2\pi}\ln z_{2}}^{\infty} \dd x    \,, \\
   &   \mathcal{B}_{3}:  \int_{-\infty}^{\infty} \dd x    \,\,\rightarrow\,\,    \int_{-\infty}^{\frac{\beta}{2\pi}\ln \bar{z}_{3}} \dd x    \,, \qquad 
        \mathcal{B}_{4}:  \int_{-\infty}^{\infty} \dd x    \,\,\rightarrow\,\,    \int_{-\infty}^{\frac{\beta}{2\pi}\ln \bar{z}_{4}} \dd x     \,.
\end{split}
\end{equation}
Then, what we only have to do is the integration over $x$ with \eqref{xRange}.
In conclusion, \eqref{RenyiInt} reads
\begin{equation} \label{RenyiResult}
\begin{split}
\int_{\mathcal{M}} \dd^{2} w  &\,  \frac{z^2 (z_{1} - z_{2})^{2}}{(z - z_{1})^2(z-z_{2})^2} \frac{\bar{z}^2 (\bar{z}_{3} - \bar{z}_{4})^{2}}{(\bar{z} - \bar{z}_{3})^2(\bar{z}-\bar{z}_{4})^2}   \\
& = \frac{\beta^2}{4 \pi }
\Biggl(
\frac{z_{1}^2 + \bar{z}_{3}^2}{(z_{1}-\bar{z}_{3})^2}   +  \frac{z_{1}^2+\bar{z}_{4}^2}{(z_{1}-\bar{z}_{4})^2}  +   \frac{z_{2}^2+\bar{z}_{3}^2}{(z_{2}-\bar{z}_{3})^2} +   \frac{z_{2}^2+\bar{z}_{4}^2}{(z_{2}-\bar{z}_{4})^2}     -    4  \\
& \qquad\qquad +  2\frac{(z_{1}+z_{2}) (\bar{z}_{3}+\bar{z}_{4})}{(z_{1}-z_{2}) (\bar{z}_{3}-\bar{z}_{4})}\ln \frac{(z_{1}-\bar{z}_{4}) (z_{2}-\bar{z}_{3})}{(z_{1}-\bar{z}_{3}) (z_{2}-\bar{z}_{4})}   
\Biggr) \,.
\end{split}
\end{equation}
%



\begin{thebibliography}{10}

\bibitem{Maldacena:1997re}
J.~M. Maldacena, \emph{{The Large N limit of superconformal field theories and
  supergravity}}, \href{http://dx.doi.org/10.1023/A:1026654312961,
  10.1023/A:1026654312961}{\emph{Adv.Theor.Math.Phys.} {\bf 2} (1998)
  231--252}, [\href{http://arxiv.org/abs/hep-th/9711200}{{\tt
  hep-th/9711200}}].

\bibitem{Gubser:1998bc}
S.~S. Gubser, I.~R. Klebanov and A.~M. Polyakov, \emph{{Gauge theory
  correlators from non-critical string theory}},
  \href{http://dx.doi.org/10.1016/S0370-2693(98)00377-3}{\emph{Phys. Lett.}
  {\bf B428} (1998) 105--114}, [\href{http://arxiv.org/abs/hep-th/9802109}{{\tt
  hep-th/9802109}}].

\bibitem{Witten:1998qj}
E.~Witten, \emph{{Anti-de Sitter space and holography}}, {\emph{Adv. Theor.
  Math. Phys.} {\bf 2} (1998) 253--291},
  [\href{http://arxiv.org/abs/hep-th/9802150}{{\tt hep-th/9802150}}].

\bibitem{McGough:2016lol}
L.~McGough, M.~Mezei and H.~Verlinde, \emph{{Moving the CFT into the bulk with
  $ T\overline{T} $}},
  \href{http://dx.doi.org/10.1007/JHEP04(2018)010}{\emph{JHEP} {\bf 04} (2018)
  010}, [\href{http://arxiv.org/abs/1611.03470}{{\tt 1611.03470}}].

\bibitem{Zamolodchikov:2004ce}
A.~B. Zamolodchikov, \emph{{Expectation value of composite field T anti-T in
  two-dimensional quantum field theory}},
  \href{http://arxiv.org/abs/hep-th/0401146}{{\tt hep-th/0401146}}.

\bibitem{Smirnov:2016lqw}
F.~A. Smirnov and A.~B. Zamolodchikov, \emph{{On space of integrable quantum
  field theories}},
  \href{http://dx.doi.org/10.1016/j.nuclphysb.2016.12.014}{\emph{Nucl. Phys.}
  {\bf B915} (2017) 363--383}, [\href{http://arxiv.org/abs/1608.05499}{{\tt
  1608.05499}}].

\bibitem{Cavaglia:2016oda}
A.~Cavagli{\`a}, S.~Negro, I.~M. Sz{\'e}cs{\'e}nyi and R.~Tateo, \emph{{$T
  \bar{T}$-deformed 2D Quantum Field Theories}},
  \href{http://dx.doi.org/10.1007/JHEP10(2016)112}{\emph{JHEP} {\bf 10} (2016)
  112}, [\href{http://arxiv.org/abs/1608.05534}{{\tt 1608.05534}}].

\bibitem{Cardy:2015xaa}
J.~Cardy, \emph{{Quantum Quenches to a Critical Point in One Dimension: some
  further results}},
  \href{http://dx.doi.org/10.1088/1742-5468/2016/02/023103}{\emph{J. Stat.
  Mech.} {\bf 1602} (2016) 023103},
  [\href{http://arxiv.org/abs/1507.07266}{{\tt 1507.07266}}].

\bibitem{Giribet:2017imm}
G.~Giribet, \emph{{$T\bar{T}$-deformations, AdS/CFT and correlation
  functions}}, \href{http://dx.doi.org/10.1007/JHEP02(2018)114}{\emph{JHEP}
  {\bf 02} (2018) 114}, [\href{http://arxiv.org/abs/1711.02716}{{\tt
  1711.02716}}].

\bibitem{Dubovsky:2017cnj}
S.~Dubovsky, V.~Gorbenko and M.~Mirbabayi, \emph{{Asymptotic fragility, near
  AdS$_{2}$ holography and $ T\overline{T} $}},
  \href{http://dx.doi.org/10.1007/JHEP09(2017)136}{\emph{JHEP} {\bf 09} (2017)
  136}, [\href{http://arxiv.org/abs/1706.06604}{{\tt 1706.06604}}].

\bibitem{Cardy:2018sdv}
J.~Cardy, \emph{{The $ T\overline{T} $ deformation of quantum field theory as
  random geometry}},
  \href{http://dx.doi.org/10.1007/JHEP10(2018)186}{\emph{JHEP} {\bf 10} (2018)
  186}, [\href{http://arxiv.org/abs/1801.06895}{{\tt 1801.06895}}].

\bibitem{Aharony:2018vux}
O.~Aharony and T.~Vaknin, \emph{{The TT* deformation at large central charge}},
  \href{http://dx.doi.org/10.1007/JHEP05(2018)166}{\emph{JHEP} {\bf 05} (2018)
  166}, [\href{http://arxiv.org/abs/1803.00100}{{\tt 1803.00100}}].

\bibitem{Bonelli:2018kik}
G.~Bonelli, N.~Doroud and M.~Zhu, \emph{{$T \bar{T}$-deformations in closed
  form}}, \href{http://dx.doi.org/10.1007/JHEP06(2018)149}{\emph{JHEP} {\bf 06}
  (2018) 149}, [\href{http://arxiv.org/abs/1804.10967}{{\tt 1804.10967}}].

\bibitem{Dubovsky:2018bmo}
S.~Dubovsky, V.~Gorbenko and G.~Hern{\'a}ndez-Chifflet, \emph{{$ T\overline{T}
  $ partition function from topological gravity}},
  \href{http://dx.doi.org/10.1007/JHEP09(2018)158}{\emph{JHEP} {\bf 09} (2018)
  158}, [\href{http://arxiv.org/abs/1805.07386}{{\tt 1805.07386}}].

\bibitem{Datta:2018thy}
S.~Datta and Y.~Jiang, \emph{{$T\bar{T}$ deformed partition functions}},
  \href{http://dx.doi.org/10.1007/JHEP08(2018)106}{\emph{JHEP} {\bf 08} (2018)
  106}, [\href{http://arxiv.org/abs/1806.07426}{{\tt 1806.07426}}].

\bibitem{Conti:2018jho}
R.~Conti, L.~Iannella, S.~Negro and R.~Tateo, \emph{{Generalised Born-Infeld
  models, Lax operators and the $ \mathrm{T}\overline{\mathrm{T}} $
  perturbation}}, \href{http://dx.doi.org/10.1007/JHEP11(2018)007}{\emph{JHEP}
  {\bf 11} (2018) 007}, [\href{http://arxiv.org/abs/1806.11515}{{\tt
  1806.11515}}].

\bibitem{Chen:2018keo}
C.~Chen, P.~Conkey, S.~Dubovsky and G.~Hern{\'a}ndez-Chifflet,
  \emph{{Undressing Confining Flux Tubes with $T\bar T$}},
  \href{http://dx.doi.org/10.1103/PhysRevD.98.114024}{\emph{Phys. Rev.} {\bf
  D98} (2018) 114024}, [\href{http://arxiv.org/abs/1808.01339}{{\tt
  1808.01339}}].

\bibitem{Aharony:2018bad}
O.~Aharony, S.~Datta, A.~Giveon, Y.~Jiang and D.~Kutasov, \emph{{Modular
  invariance and uniqueness of $T\bar{T}$ deformed CFT}},
  \href{http://dx.doi.org/10.1007/JHEP01(2019)086}{\emph{JHEP} {\bf 01} (2019)
  086}, [\href{http://arxiv.org/abs/1808.02492}{{\tt 1808.02492}}].

\bibitem{Conti:2018tca}
R.~Conti, S.~Negro and R.~Tateo, \emph{{The $ \mathrm{T}\overline{\mathrm{T}} $
  perturbation and its geometric interpretation}},
  \href{http://dx.doi.org/10.1007/JHEP02(2019)085}{\emph{JHEP} {\bf 02} (2019)
  085}, [\href{http://arxiv.org/abs/1809.09593}{{\tt 1809.09593}}].

\bibitem{Santilli:2018xux}
L.~Santilli and M.~Tierz, \emph{{Large N phase transition in $ T\overline{T} $
  -deformed 2d Yang-Mills theory on the sphere}},
  \href{http://dx.doi.org/10.1007/JHEP01(2019)054}{\emph{JHEP} {\bf 01} (2019)
  054}, [\href{http://arxiv.org/abs/1810.05404}{{\tt 1810.05404}}].

\bibitem{Baggio:2018rpv}
M.~Baggio, A.~Sfondrini, G.~Tartaglino-Mazzucchelli and H.~Walsh, \emph{{On $
  T\overline{T} $ deformations and supersymmetry}},
  \href{http://dx.doi.org/10.1007/JHEP06(2019)063}{\emph{JHEP} {\bf 06} (2019)
  063}, [\href{http://arxiv.org/abs/1811.00533}{{\tt 1811.00533}}].

\bibitem{Chang:2018dge}
C.-K. Chang, C.~Ferko and S.~Sethi, \emph{{Supersymmetry and $ T\overline{T} $
  deformations}}, \href{http://dx.doi.org/10.1007/JHEP04(2019)131}{\emph{JHEP}
  {\bf 04} (2019) 131}, [\href{http://arxiv.org/abs/1811.01895}{{\tt
  1811.01895}}].

\bibitem{Jiang:2019tcq}
Y.~Jiang, \emph{{Expectation value of $\mathrm{T}\overline{\mathrm{T}}$
  operator in curved spacetimes}},  \href{http://arxiv.org/abs/1903.07561}{{\tt
  1903.07561}}.

\bibitem{LeFloch:2019rut}
B.~Le~Floch and M.~Mezei, \emph{{Solving a family of $T\bar{T}$-like
  theories}},  \href{http://arxiv.org/abs/1903.07606}{{\tt 1903.07606}}.

\bibitem{Jiang:2019hux}
H.~Jiang, A.~Sfondrini and G.~Tartaglino-Mazzucchelli, \emph{{$T\bar{T}$
  deformations with $\mathcal{N}=(0,2)$ supersymmetry}},
  \href{http://dx.doi.org/10.1103/PhysRevD.100.046017}{\emph{Phys. Rev.} {\bf
  D100} (2019) 046017}, [\href{http://arxiv.org/abs/1904.04760}{{\tt
  1904.04760}}].

\bibitem{Conti:2019dxg}
R.~Conti, S.~Negro and R.~Tateo, \emph{{Conserved currents and
  $\text{T}\bar{\text{T}}_s$ irrelevant deformations of 2D integrable field
  theories}}, \href{http://dx.doi.org/10.1007/JHEP11(2019)120}{\emph{JHEP} {\bf
  11} (2019) 120}, [\href{http://arxiv.org/abs/1904.09141}{{\tt 1904.09141}}].

\bibitem{Chang:2019kiu}
C.-K. Chang, C.~Ferko, S.~Sethi, A.~Sfondrini and G.~Tartaglino-Mazzucchelli,
  \emph{{$T\bar{T}$ flows and (2,2) supersymmetry}},
  \href{http://dx.doi.org/10.1103/PhysRevD.101.026008}{\emph{Phys. Rev.} {\bf
  D101} (2020) 026008}, [\href{http://arxiv.org/abs/1906.00467}{{\tt
  1906.00467}}].

\bibitem{Frolov:2019nrr}
S.~Frolov, \emph{{TTbar deformation and the light-cone gauge}},
  \href{http://arxiv.org/abs/1905.07946}{{\tt 1905.07946}}.

\bibitem{Guica:2017lia}
M.~Guica, \emph{{An integrable Lorentz-breaking deformation of two-dimensional
  CFTs}}, \href{http://dx.doi.org/10.21468/SciPostPhys.5.5.048}{\emph{SciPost
  Phys.} {\bf 5} (2018) 048}, [\href{http://arxiv.org/abs/1710.08415}{{\tt
  1710.08415}}].

\bibitem{Chakraborty:2018vja}
S.~Chakraborty, A.~Giveon and D.~Kutasov, \emph{{$ J\overline{T} $ deformed
  CFT$_{2}$ and string theory}},
  \href{http://dx.doi.org/10.1007/JHEP10(2018)057}{\emph{JHEP} {\bf 10} (2018)
  057}, [\href{http://arxiv.org/abs/1806.09667}{{\tt 1806.09667}}].

\bibitem{Aharony:2018ics}
O.~Aharony, S.~Datta, A.~Giveon, Y.~Jiang and D.~Kutasov, \emph{{Modular
  covariance and uniqueness of $J\bar{T}$ deformed CFTs}},
  \href{http://dx.doi.org/10.1007/JHEP01(2019)085}{\emph{JHEP} {\bf 01} (2019)
  085}, [\href{http://arxiv.org/abs/1808.08978}{{\tt 1808.08978}}].

\bibitem{Cardy:2018jho}
J.~Cardy, \emph{{$T\overline T$ deformations of non-Lorentz invariant field
  theories}},  \href{http://arxiv.org/abs/1809.07849}{{\tt 1809.07849}}.

\bibitem{Nakayama:2018ujt}
Y.~Nakayama, \emph{{Very Special $T\bar{J}$ deformed CFT}},
  \href{http://dx.doi.org/10.1103/PhysRevD.99.085008}{\emph{Phys. Rev.} {\bf
  D99} (2019) 085008}, [\href{http://arxiv.org/abs/1811.02173}{{\tt
  1811.02173}}].

\bibitem{Guica:2019vnb}
M.~Guica, \emph{{On correlation functions in $J\bar T$-deformed CFTs}},
  \href{http://dx.doi.org/10.1088/1751-8121/ab0ef3}{\emph{J. Phys.} {\bf A52}
  (2019) 184003}, [\href{http://arxiv.org/abs/1902.01434}{{\tt 1902.01434}}].

\bibitem{Shyam:2017znq}
V.~Shyam, \emph{{Background independent holographic dual to $T\bar{T}$ deformed
  CFT with large central charge in 2 dimensions}},
  \href{http://dx.doi.org/10.1007/JHEP10(2017)108}{\emph{JHEP} {\bf 10} (2017)
  108}, [\href{http://arxiv.org/abs/1707.08118}{{\tt 1707.08118}}].

\bibitem{Kraus:2018xrn}
P.~Kraus, J.~Liu and D.~Marolf, \emph{{Cutoff AdS$_{3}$ versus the $
  T\overline{T} $ deformation}},
  \href{http://dx.doi.org/10.1007/JHEP07(2018)027}{\emph{JHEP} {\bf 07} (2018)
  027}, [\href{http://arxiv.org/abs/1801.02714}{{\tt 1801.02714}}].

\bibitem{Cottrell:2018skz}
W.~Cottrell and A.~Hashimoto, \emph{{Comments on $T \bar T$ double trace
  deformations and boundary conditions}},
  \href{http://dx.doi.org/10.1016/j.physletb.2018.09.068}{\emph{Phys. Lett.}
  {\bf B789} (2019) 251--255}, [\href{http://arxiv.org/abs/1801.09708}{{\tt
  1801.09708}}].

\bibitem{Bzowski:2018pcy}
A.~Bzowski and M.~Guica, \emph{{The holographic interpretation of $J \bar
  T$-deformed CFTs}},
  \href{http://dx.doi.org/10.1007/JHEP01(2019)198}{\emph{JHEP} {\bf 01} (2019)
  198}, [\href{http://arxiv.org/abs/1803.09753}{{\tt 1803.09753}}].

\bibitem{Taylor:2018xcy}
M.~Taylor, \emph{{TT deformations in general dimensions}},
  \href{http://arxiv.org/abs/1805.10287}{{\tt 1805.10287}}.

\bibitem{Hartman:2018tkw}
T.~Hartman, J.~Kruthoff, E.~Shaghoulian and A.~Tajdini, \emph{{Holography at
  finite cutoff with a $T^2$ deformation}},
  \href{http://dx.doi.org/10.1007/JHEP03(2019)004}{\emph{JHEP} {\bf 03} (2019)
  004}, [\href{http://arxiv.org/abs/1807.11401}{{\tt 1807.11401}}].

\bibitem{Shyam:2018sro}
V.~Shyam, \emph{{Finite Cutoff AdS$_{5}$ Holography and the Generalized
  Gradient Flow}}, \href{http://dx.doi.org/10.1007/JHEP12(2018)086}{\emph{JHEP}
  {\bf 12} (2018) 086}, [\href{http://arxiv.org/abs/1808.07760}{{\tt
  1808.07760}}].

\bibitem{Caputa:2019pam}
P.~Caputa, S.~Datta and V.~Shyam, \emph{{Sphere partition functions \& cut-off
  AdS}}, \href{http://dx.doi.org/10.1007/JHEP05(2019)112}{\emph{JHEP} {\bf 05}
  (2019) 112}, [\href{http://arxiv.org/abs/1902.10893}{{\tt 1902.10893}}].

\bibitem{Giveon:2017nie}
A.~Giveon, N.~Itzhaki and D.~Kutasov, \emph{{$ \mathrm{T}\overline{\mathrm{T}}
  $ and LST}}, \href{http://dx.doi.org/10.1007/JHEP07(2017)122}{\emph{JHEP}
  {\bf 07} (2017) 122}, [\href{http://arxiv.org/abs/1701.05576}{{\tt
  1701.05576}}].

\bibitem{Giveon:2017myj}
A.~Giveon, N.~Itzhaki and D.~Kutasov, \emph{{A solvable irrelevant deformation
  of AdS$_{3}$/CFT$_{2}$}},
  \href{http://dx.doi.org/10.1007/JHEP12(2017)155}{\emph{JHEP} {\bf 12} (2017)
  155}, [\href{http://arxiv.org/abs/1707.05800}{{\tt 1707.05800}}].

\bibitem{Asrat:2017tzd}
M.~Asrat, A.~Giveon, N.~Itzhaki and D.~Kutasov, \emph{{Holography Beyond AdS}},
  \href{http://dx.doi.org/10.1016/j.nuclphysb.2018.05.005}{\emph{Nucl. Phys.}
  {\bf B932} (2018) 241--253}, [\href{http://arxiv.org/abs/1711.02690}{{\tt
  1711.02690}}].

\bibitem{Baggio:2018gct}
M.~Baggio and A.~Sfondrini, \emph{{Strings on NS-NS Backgrounds as Integrable
  Deformations}},
  \href{http://dx.doi.org/10.1103/PhysRevD.98.021902}{\emph{Phys. Rev.} {\bf
  D98} (2018) 021902}, [\href{http://arxiv.org/abs/1804.01998}{{\tt
  1804.01998}}].

\bibitem{Apolo:2018qpq}
L.~Apolo and W.~Song, \emph{{Strings on warped AdS$_{3}$ via $
  \mathrm{T}\bar{\mathrm{J}} $ deformations}},
  \href{http://dx.doi.org/10.1007/JHEP10(2018)165}{\emph{JHEP} {\bf 10} (2018)
  165}, [\href{http://arxiv.org/abs/1806.10127}{{\tt 1806.10127}}].

\bibitem{Babaro:2018cmq}
J.~P. Babaro, V.~F. Foit, G.~Giribet and M.~Leoni, \emph{{$ T\overline{T} $
  type deformation in the presence of a boundary}},
  \href{http://dx.doi.org/10.1007/JHEP08(2018)096}{\emph{JHEP} {\bf 08} (2018)
  096}, [\href{http://arxiv.org/abs/1806.10713}{{\tt 1806.10713}}].

\bibitem{Chakraborty:2018aji}
S.~Chakraborty, \emph{{Wilson loop in a $T\bar{T}$ like deformed
  $\rm{CFT}_2$}},
  \href{http://dx.doi.org/10.1016/j.nuclphysb.2018.12.003}{\emph{Nucl. Phys.}
  {\bf B938} (2019) 605--620}, [\href{http://arxiv.org/abs/1809.01915}{{\tt
  1809.01915}}].

\bibitem{Araujo:2018rho}
T.~Araujo, E.~Colg\'{a}in, Y.~Sakatani, M.~M. Sheikh-Jabbari and H.~Yavartanoo,
  \emph{{Holographic integration of $T \bar{T}$ \& $J \bar{T}$ via $O(d,d)$}},
  \href{http://dx.doi.org/10.1007/JHEP03(2019)168}{\emph{JHEP} {\bf 03} (2019)
  168}, [\href{http://arxiv.org/abs/1811.03050}{{\tt 1811.03050}}].

\bibitem{Giveon:2019fgr}
A.~Giveon, \emph{{Comments on $T\bar T$, $J\bar{T}$ and String Theory}},
  \href{http://arxiv.org/abs/1903.06883}{{\tt 1903.06883}}.

\bibitem{Chakraborty:2019mdf}
S.~Chakraborty, A.~Giveon and D.~Kutasov, \emph{{$T\bar{T}$, $J\bar{T}$,
  $T\bar{J}$ and String Theory}},
  \href{http://dx.doi.org/10.1088/1751-8121/ab3710}{\emph{J. Phys.} {\bf A52}
  (2019) 384003}, [\href{http://arxiv.org/abs/1905.00051}{{\tt 1905.00051}}].

\bibitem{Nakayama:2019mvq}
Y.~Nakayama, \emph{{Holographic dual of conformal field theories with very
  special $T\bar{J}$ deformations}},
  \href{http://dx.doi.org/10.1103/PhysRevD.100.086011}{\emph{Phys. Rev.} {\bf
  D100} (2019) 086011}, [\href{http://arxiv.org/abs/1905.05353}{{\tt
  1905.05353}}].

\bibitem{Dei:2018mfl}
A.~Dei and A.~Sfondrini, \emph{{Integrable spin chain for stringy
  Wess-Zumino-Witten models}},
  \href{http://dx.doi.org/10.1007/JHEP07(2018)109}{\emph{JHEP} {\bf 07} (2018)
  109}, [\href{http://arxiv.org/abs/1806.00422}{{\tt 1806.00422}}].

\bibitem{Dei:2018jyj}
A.~Dei and A.~Sfondrini, \emph{{Integrable S matrix, mirror TBA and spectrum
  for the stringy AdS$_{3}$ $\times$ S$^{3}$ $\times$ S$^{3}$ $\times$ S$^{1}$
  WZW model}}, \href{http://dx.doi.org/10.1007/JHEP02(2019)072}{\emph{JHEP}
  {\bf 02} (2019) 072}, [\href{http://arxiv.org/abs/1812.08195}{{\tt
  1812.08195}}].

\bibitem{Ryu:2006bv}
S.~Ryu and T.~Takayanagi, \emph{{Holographic derivation of entanglement entropy
  from AdS/CFT}},
  \href{http://dx.doi.org/10.1103/PhysRevLett.96.181602}{\emph{Phys. Rev.
  Lett.} {\bf 96} (2006) 181602},
  [\href{http://arxiv.org/abs/hep-th/0603001}{{\tt hep-th/0603001}}].

\bibitem{Ryu:2006ef}
S.~Ryu and T.~Takayanagi, \emph{{Aspects of Holographic Entanglement Entropy}},
  \href{http://dx.doi.org/10.1088/1126-6708/2006/08/045}{\emph{JHEP} {\bf 08}
  (2006) 045}, [\href{http://arxiv.org/abs/hep-th/0605073}{{\tt
  hep-th/0605073}}].

\bibitem{Chakraborty:2018kpr}
S.~Chakraborty, A.~Giveon, N.~Itzhaki and D.~Kutasov, \emph{{Entanglement
  beyond AdS}},
  \href{http://dx.doi.org/10.1016/j.nuclphysb.2018.08.011}{\emph{Nucl. Phys.}
  {\bf B935} (2018) 290--309}, [\href{http://arxiv.org/abs/1805.06286}{{\tt
  1805.06286}}].

\bibitem{Donnelly:2018bef}
W.~Donnelly and V.~Shyam, \emph{{Entanglement entropy and $T \overline{T}$
  deformation}},
  \href{http://dx.doi.org/10.1103/PhysRevLett.121.131602}{\emph{Phys. Rev.
  Lett.} {\bf 121} (2018) 131602}, [\href{http://arxiv.org/abs/1806.07444}{{\tt
  1806.07444}}].

\bibitem{Chen:2018eqk}
B.~Chen, L.~Chen and P.-X. Hao, \emph{{Entanglement entropy in
  $T\overline{T}$-deformed CFT}},
  \href{http://dx.doi.org/10.1103/PhysRevD.98.086025}{\emph{Phys. Rev.} {\bf
  D98} (2018) 086025}, [\href{http://arxiv.org/abs/1807.08293}{{\tt
  1807.08293}}].

\bibitem{Gorbenko:2018oov}
V.~Gorbenko, E.~Silverstein and G.~Torroba, \emph{{dS/dS and $ T\overline{T}
  $}}, \href{http://dx.doi.org/10.1007/JHEP03(2019)085}{\emph{JHEP} {\bf 03}
  (2019) 085}, [\href{http://arxiv.org/abs/1811.07965}{{\tt 1811.07965}}].

\bibitem{Park:2018snf}
C.~Park, \emph{{Holographic Entanglement Entropy in Cutoff AdS}},
  \href{http://dx.doi.org/10.1142/S0217751X18502263}{\emph{Int. J. Mod. Phys.}
  {\bf A33} (2019) 1850226}, [\href{http://arxiv.org/abs/1812.00545}{{\tt
  1812.00545}}].

\bibitem{Sun:2019ijq}
Y.~Sun and J.-R. Sun, \emph{{Note on R\'{e}nyi entropy of 2D perturbed free
  fermions}}, \href{http://dx.doi.org/10.1103/PhysRevD.99.106008}{\emph{Phys.
  Rev.} {\bf D99} (2019) 106008}, [\href{http://arxiv.org/abs/1901.08796}{{\tt
  1901.08796}}].

\bibitem{Banerjee:2019ewu}
A.~Banerjee, A.~Bhattacharyya and S.~Chakraborty, \emph{{Entanglement Entropy
  for $TT$ deformed CFT in general dimensions}},
  \href{http://dx.doi.org/10.1016/j.nuclphysb.2019.114775}{\emph{Nucl. Phys.}
  {\bf B948} (2019) 114775}, [\href{http://arxiv.org/abs/1904.00716}{{\tt
  1904.00716}}].

\bibitem{Murdia:2019fax}
C.~Murdia, Y.~Nomura, P.~Rath and N.~Salzetta, \emph{{Comments on holographic
  entanglement entropy in $TT$ deformed conformal field theories}},
  \href{http://dx.doi.org/10.1103/PhysRevD.100.026011}{\emph{Phys. Rev.} {\bf
  D100} (2019) 026011}, [\href{http://arxiv.org/abs/1904.04408}{{\tt
  1904.04408}}].

\bibitem{Ota:2019yfe}
T.~Ota, \emph{{Comments on holographic entanglements in cutoff AdS}},
  \href{http://arxiv.org/abs/1904.06930}{{\tt 1904.06930}}.

\bibitem{Calabrese:2004eu}
P.~Calabrese and J.~L. Cardy, \emph{{Entanglement entropy and quantum field
  theory}}, \href{http://dx.doi.org/10.1088/1742-5468/2004/06/P06002}{\emph{J.
  Stat. Mech.} {\bf 0406} (2004) P06002},
  [\href{http://arxiv.org/abs/hep-th/0405152}{{\tt hep-th/0405152}}].

\bibitem{Calabrese:2009qy}
P.~Calabrese and J.~Cardy, \emph{{Entanglement entropy and conformal field
  theory}}, \href{http://dx.doi.org/10.1088/1751-8113/42/50/504005}{\emph{J.
  Phys.} {\bf A42} (2009) 504005}, [\href{http://arxiv.org/abs/0905.4013}{{\tt
  0905.4013}}].

\bibitem{Hartman:2013mia}
T.~Hartman, \emph{{Entanglement Entropy at Large Central Charge}},
  \href{http://arxiv.org/abs/1303.6955}{{\tt 1303.6955}}.

\bibitem{Cardy:2007mb}
J.~L. Cardy, O.~A. Castro-Alvaredo and B.~Doyon, \emph{{Form factors of
  branch-point twist fields in quantum integrable models and entanglement
  entropy}}, \href{http://dx.doi.org/10.1007/s10955-007-9422-x}{\emph{J.
  Statist. Phys.} {\bf 130} (2008) 129--168},
  [\href{http://arxiv.org/abs/0706.3384}{{\tt 0706.3384}}].

\bibitem{Fitzpatrick:2014vua}
A.~L. Fitzpatrick, J.~Kaplan and M.~T. Walters, \emph{{Universality of
  Long-Distance AdS Physics from the CFT Bootstrap}},
  \href{http://dx.doi.org/10.1007/JHEP08(2014)145}{\emph{JHEP} {\bf 08} (2014)
  145}, [\href{http://arxiv.org/abs/1403.6829}{{\tt 1403.6829}}].

\bibitem{Perlmutter:2015iya}
E.~Perlmutter, \emph{{Virasoro conformal blocks in closed form}},
  \href{http://dx.doi.org/10.1007/JHEP08(2015)088}{\emph{JHEP} {\bf 08} (2015)
  088}, [\href{http://arxiv.org/abs/1502.07742}{{\tt 1502.07742}}].

\bibitem{DiFrancesco:1997nk}
P.~Di~Francesco, P.~Mathieu and D.~Senechal, \emph{{Conformal Field Theory}}.
\newblock Graduate Texts in Contemporary Physics. Springer-Verlag, New York,
  1997,
  \href{http://dx.doi.org/10.1007/978-1-4612-2256-9}{10.1007/978-1-4612-2256-9}.

\bibitem{Hartman:2015lfa}
T.~Hartman, S.~Jain and S.~Kundu, \emph{{Causality Constraints in Conformal
  Field Theory}}, \href{http://dx.doi.org/10.1007/JHEP05(2016)099}{\emph{JHEP}
  {\bf 05} (2016) 099}, [\href{http://arxiv.org/abs/1509.00014}{{\tt
  1509.00014}}].

\bibitem{Headrick:2010zt}
M.~Headrick, \emph{{Entanglement Renyi entropies in holographic theories}},
  \href{http://dx.doi.org/10.1103/PhysRevD.82.126010}{\emph{Phys. Rev.} {\bf
  D82} (2010) 126010}, [\href{http://arxiv.org/abs/1006.0047}{{\tt
  1006.0047}}].

\bibitem{Calabrese:2009ez}
P.~Calabrese, J.~Cardy and E.~Tonni, \emph{{Entanglement entropy of two
  disjoint intervals in conformal field theory}},
  \href{http://dx.doi.org/10.1088/1742-5468/2009/11/P11001}{\emph{J. Stat.
  Mech.} {\bf 0911} (2009) P11001}, [\href{http://arxiv.org/abs/0905.2069}{{\tt
  0905.2069}}].

\bibitem{Dubovsky:2012wk}
S.~Dubovsky, R.~Flauger and V.~Gorbenko, \emph{{Solving the Simplest Theory of
  Quantum Gravity}},
  \href{http://dx.doi.org/10.1007/JHEP09(2012)133}{\emph{JHEP} {\bf 09} (2012)
  133}, [\href{http://arxiv.org/abs/1205.6805}{{\tt 1205.6805}}].

\bibitem{Dubovsky:2013ira}
S.~Dubovsky, V.~Gorbenko and M.~Mirbabayi, \emph{{Natural Tuning: Towards A
  Proof of Concept}},
  \href{http://dx.doi.org/10.1007/JHEP09(2013)045}{\emph{JHEP} {\bf 09} (2013)
  045}, [\href{http://arxiv.org/abs/1305.6939}{{\tt 1305.6939}}].

\bibitem{Cooper:2013ffa}
P.~Cooper, S.~Dubovsky and A.~Mohsen, \emph{{Ultraviolet complete
  Lorentz-invariant theory with superluminal signal propagation}},
  \href{http://dx.doi.org/10.1103/PhysRevD.89.084044}{\emph{Phys. Rev.} {\bf
  D89} (2014) 084044}, [\href{http://arxiv.org/abs/1312.2021}{{\tt
  1312.2021}}].

\bibitem{Dong:2016fnf}
X.~Dong, \emph{{The Gravity Dual of Renyi Entropy}},
  \href{http://dx.doi.org/10.1038/ncomms12472}{\emph{Nature Commun.} {\bf 7}
  (2016) 12472}, [\href{http://arxiv.org/abs/1601.06788}{{\tt 1601.06788}}].

\bibitem{Strominger:1997eq}
A.~Strominger, \emph{{Black hole entropy from near horizon microstates}},
  \href{http://dx.doi.org/10.1088/1126-6708/1998/02/009}{\emph{JHEP} {\bf 02}
  (1998) 009}, [\href{http://arxiv.org/abs/hep-th/9712251}{{\tt
  hep-th/9712251}}].

\bibitem{Fischler:2012uv}
W.~Fischler, A.~Kundu and S.~Kundu, \emph{{Holographic Mutual Information at
  Finite Temperature}},
  \href{http://dx.doi.org/10.1103/PhysRevD.87.126012}{\emph{Phys. Rev.} {\bf
  D87} (2013) 126012}, [\href{http://arxiv.org/abs/1212.4764}{{\tt
  1212.4764}}].

\bibitem{BabaeiVelni:2019pkw}
K.~Babaei~Velni, M.~R. Mohammadi~Mozaffar and M.~H. Vahidinia, \emph{{Some
  Aspects of Entanglement Wedge Cross-Section}},
  \href{http://dx.doi.org/10.1007/JHEP05(2019)200}{\emph{JHEP} {\bf 05} (2019)
  200}, [\href{http://arxiv.org/abs/1903.08490}{{\tt 1903.08490}}].

\bibitem{Gubser:2008px}
S.~S. Gubser, \emph{{Breaking an Abelian gauge symmetry near a black hole
  horizon}}, \href{http://dx.doi.org/10.1103/PhysRevD.78.065034}{\emph{Phys.
  Rev.} {\bf D78} (2008) 065034}, [\href{http://arxiv.org/abs/0801.2977}{{\tt
  0801.2977}}].

\bibitem{Hartnoll:2008vx}
S.~A. Hartnoll, C.~P. Herzog and G.~T. Horowitz, \emph{{Building a Holographic
  Superconductor}},
  \href{http://dx.doi.org/10.1103/PhysRevLett.101.031601}{\emph{Phys.Rev.Lett.}
  {\bf 101} (2008) 031601}, [\href{http://arxiv.org/abs/0803.3295}{{\tt
  0803.3295}}].

\bibitem{Calabrese:2012ew}
P.~Calabrese, J.~Cardy and E.~Tonni, \emph{{Entanglement negativity in quantum
  field theory}},
  \href{http://dx.doi.org/10.1103/PhysRevLett.109.130502}{\emph{Phys. Rev.
  Lett.} {\bf 109} (2012) 130502}, [\href{http://arxiv.org/abs/1206.3092}{{\tt
  1206.3092}}].

\bibitem{Tamaoka:2018ned}
K.~Tamaoka, \emph{{Entanglement Wedge Cross Section from the Dual Density
  Matrix}}, \href{http://dx.doi.org/10.1103/PhysRevLett.122.141601}{\emph{Phys.
  Rev. Lett.} {\bf 122} (2019) 141601},
  [\href{http://arxiv.org/abs/1809.09109}{{\tt 1809.09109}}].

\bibitem{Caputa:2018xuf}
P.~Caputa, M.~Miyaji, T.~Takayanagi and K.~Umemoto, \emph{{Holographic
  Entanglement of Purification from Conformal Field Theories}},
  \href{http://dx.doi.org/10.1103/PhysRevLett.122.111601}{\emph{Phys. Rev.
  Lett.} {\bf 122} (2019) 111601}, [\href{http://arxiv.org/abs/1812.05268}{{\tt
  1812.05268}}].

\bibitem{Dutta:2019gen}
S.~Dutta and T.~Faulkner, \emph{{A canonical purification for the entanglement
  wedge cross-section}},  \href{http://arxiv.org/abs/1905.00577}{{\tt
  1905.00577}}.

\end{thebibliography}
\bibliographystyle{JHEP}
\providecommand{\href}[2]{#2}\begingroup\raggedright\endgroup

\end{document}